\documentclass[reqno, 10pt]{amsart}

\usepackage[top=4cm, bottom=3cm, left=3cm, right=3cm]{geometry}
\usepackage{amsmath}

\newtheorem{thm}{Theorem}
\newtheorem{remark}{Remark}

\newtheorem{lem}{Lemma}
\newtheorem{prop}{Proposition}
\theoremstyle{definition}

\theoremstyle{remark}

\newcommand{\U}{\mathcal{U}}
\newcommand{\N}{\mathcal{N}}
\newcommand{\h}{\mathcal{H}}
\newcommand{\F}{\mathcal{F}}
\newcommand{\R}{\mathcal{R}}
\newcommand{\A}{\mathcal{A}}
\newcommand{\T}{\mathcal{T}}
\newcommand{\D}{\mathcal{D}}
\newcommand{\Q}{\mathcal{Q}}

\numberwithin{equation}{section} \numberwithin{lem}{section}
\numberwithin{thm}{section} \numberwithin{prop}{section}
\numberwithin{cor}{section} \numberwithin{rem}{section}

\begin{document}

\thispagestyle{headings}

\title{Rate of Convergence in Nonlinear Hartree Dynamics with Factorized Initial Data}

\author{Li Chen}
\address{Department of Mathematical Sciences, Tsinghua University, Beijing, 100084, People's Republic of China} \email{{\tt lchen@math.tsinghua.edu.cn}}
\thanks{Li Chen is partially supported by National Natural Science Foundation of China (NSFC), grant number 10871112}

\author{Ji Oon Lee}
\address{Department of Mathematical Sciences, Korea Advanced Institute of Science and Technology, Daejeon, 305701, Republic of Korea}
\email{jioon.lee@kaist.edu}


\begin{abstract}
The mean field dynamics of an $N$-particle weekly interacting Boson system can be described by the nonlinear Hartree equation. In this paper, we present estimates on the $1/N$ rate of convergence of many-body Schr\"{o}dinger dynamics to the one-body nonlinear Hartree dynamics with factorized initial data with two-body interaction potential $V$ in $L^3 (\mathbb{R}^3)+ L^{\infty} (\mathbb{R}^3)$.
\end{abstract}

\maketitle

\section{Introduction and Main Result}

\subsection{Setting and history of the problem}

In a non-relativistic case, the dynamics of an $N$-particle system is governed by the Schr\"{o}dinger equation. For a real physical system, however, $N$ is usually very large so that it is hopeless to solve the $N$-body Schr\"{o}dinger equation directly. There were many efforts to describe such a system by approximating by a simpler dynamics. One of the most important cases is a system of $N$-weakly interacting Bosons, which can be approximated well by using the nonlinear Hartree equation.

We consider a system of $N$-interacting three-dimensional Bosons in $\mathbb{R}^3$, described on $(L^2(\mathbb{R}^{3N}))_s$, the subspace of $L^2(\mathbb{R}^{3N}, dX_N)$ consisting of all symmetric functions, where $X_N := (x_1, x_2, \cdots, x_N)$. Given the two-particle interaction $V$, the mean-field Hamiltonian of this system is 
\begin{equation}
H_N := -\sum_{j=1}^N \Delta_j + \frac{1}{N} \sum_{i<j}^N V(x_i - x_j),
\end{equation}
where $x_i \in \mathbb{R}^3$ are the positions of the particles and $\Delta_i$ denotes the Laplacian with respect to $i$-th particle. Note that the coupling constant $1/N$ guarantees that the kinetic energy and the potential energy are typically of the same order.

We assume that the initial state is factorized, i.e.
\begin{equation}
\psi(X_N) = \prod_{j=1}^N \varphi(x_j) \label{factorized initial state}
\end{equation}
for some $\varphi \in H^1 (\mathbb{R}^3)$ with $\| \varphi \|_{L^2(\mathbb{R}^3)} = 1$. Its time evolution, $\psi_t$, satisfies the $N$-body Schr\"{o}dinger equation,
\begin{equation}
i \partial_t \psi_t = H_N \psi_t
\end{equation}
with the initial data $\psi_0 = \psi$. 

It is well known in this system that the factorization is approximately preserved under the time evolution, and, in fact, we can expect that
\begin{equation}
\psi_t(X_N) \simeq \prod_{j=1}^N \varphi_t (x_j),
\label{approximately factorized}
\end{equation}
where $\varphi_t$ is given by the solution of the nonlinear Hartree equation,
\begin{equation}
i \partial_t \varphi_t = -\Delta \varphi_t + (V * |\varphi_t|^2 ) \varphi_t, \qquad \varphi_t |_{t=0} = \varphi. \label{Hartree}
\end{equation}

To give a meaning to \eqref{approximately factorized}, we introduce marginal densities. The marginal density associated with $\psi_t$ is defined to be the orthogonal projection onto $\psi_t$, and we denote this by bracket notation,
\begin{equation}
\gamma_t = | \psi_t \rangle \langle \psi_t |.
\end{equation}
The kernel of $\gamma_t$ is given by
\begin{equation}
\gamma_t (X_N; X_N') = \psi_t(X_N) \overline{\psi_t} (X_N').
\end{equation}
For $k=1, \cdots, N-1$, we also define $k$-particle marginal density $\gamma_t^{(k)}$ associated with $\psi_t$ by taking the partial trace of $\gamma_t$ over the last $(N-k)$ particles, i.e., the kernel of $\gamma_t^{(k)}$ is given by
\begin{equation}
\gamma_t^{(k)}(X_k; X_k') = \int dx_{k+1} \cdots dx_N \gamma_t (X_k, x_{k+1}, \cdots, x_N; X_k', x_{k+1}, \cdots, x_N ), \label{marginal density}
\end{equation}
where $X_k = (x_1, \cdots, x_k)$ and $X_k' = (x'_1, \cdots, x'_k)$.

In terms of marginal densities, \eqref{factorized initial state} is equivalent to
\begin{equation}
\gamma = \gamma_0 = |\varphi \rangle \langle \varphi |^{\otimes N},
\end{equation}
and it is expected to show that its time evolution satisfies the convergence
\begin{equation}
\gamma_t^{(k)} \rightarrow |\varphi_t \rangle \langle \varphi_t |^{\otimes k}  \;\;\;\; \text{as} \;\; N \rightarrow \infty
\label{convergence}
\end{equation}
in the trace norm topology.

Spohn \cite{S} proved \eqref{convergence}, assuming the interaction potential $V$ is bounded, and Erd\"{o}s and Yau \cite{EY} extended this result further to obtain a rigorous derivation of the Hartree equation for the Coulomb interaction case. These results were based on the study of dynamics of marginal densities, which is governed by BBGKY hierarchy
\begin{align}
i \partial_t \gamma_t^{(k)} &= \sum_{j=1}^k \big[ -\Delta_j, \gamma_t^{(k)} \big] + \frac{1}{N} \sum_{i<j}^k \big[ V(x_i - x_j), \gamma_t^{(k)} \big] \\
& \quad +\frac{N-k}{N} \sum_{j=1}^k \textrm{Tr}_{k+1} \big[ V(x_j - x_{k+1}), \gamma_t^{(k+1)} \big], \nonumber
\end{align}
where $\textrm{Tr}_{k+1}$ denotes the partial trace over the $(k+1)$-st particle. In this method, compactness of the sequence of marginal densities $\{ \gamma_t^{(k)} \}_{k=1}^N$ is first shown, then that any limit point of the sequence is a solution of the infinite hierarchy is proved, and finally, proving the uniqueness of the solution gives the desired result.

Another approach was introduced by Hepp \cite{H} and extended by Ginibre and Velo \cite{GV1, GV2}. In this approach, the time evolution of coherent states was studied in the second quantized Fock-space representation. Using this method, the rate of convergence in \eqref{convergence} was proved by Rodnianski and Schlein \cite{RS}, where they proved that, for factorized initial data and for the interaction potential such that $V^2 (x) \leq C(1- \Delta_x)$,
\begin{equation}
\textrm{Tr} \big| \gamma_t^{(k)} - | \varphi_t \rangle \langle \varphi_t |^{\otimes k} \big| \leq \frac{C e^{Kt}}{\sqrt N} \label{1/2 estimate}
\end{equation}
for some constants $C$ and $K$.

Knowles and Pickl \cite{KP} introduced a new method to prove \eqref{1/2 estimate} where the interaction potential $V \in L^2 (\mathbb{R}^3) + L^{\infty} (\mathbb{R}^3)$. They also extended the result to more singular interaction potentials, where the rate of convergence becomes weaker as the singularity in the interaction potential stronger.

The optimal result for this quantitative estimate must be of the order $1/N$, since the bound of the form
\begin{equation}
\textrm{Tr} \big| \gamma_t^{(k)} - | \varphi_t \rangle \langle \varphi_t |^{\otimes k} \big| \leq \frac{C e^{Kt}}{N}
\end{equation}
is proved for the coherent case by Rodnianski and Schlein \cite{RS} and for the factorized initial state case with the bounded interaction potential by Erd\"os and Schlein \cite{ES}.

In this paper, we first observe that the $\U_2$ dynamics, which was introduced by Ginibre and Velo \cite{GV1, GV2}, gives the $O(1/N)$ rate of convergence. The goal of this paper is to show that for factorized initial data, there exist constants $C$ and $K$ and $\| \varphi \|_{H^1}$ such that
\begin{equation}
\textrm{Tr} \big| \gamma_t^{(k)} - | \varphi_t \rangle \langle \varphi_t |^{\otimes k} \big| \leq \frac{C e^{Kt}}{N} \label{1/N estimate}
\end{equation}
for $V \in L^3(\mathbb{R}^3) + L^\infty(\mathbb{R}^3)$.

\subsection{Notations and tools to be used}

Let $\F_X$ be the Fock space of symmetric functions, i.e.
\begin{equation}
\F_X := \bigoplus_{n \geq 0} \big( L^2 (\mathbb{R}^{3n}) \big)_s,
\end{equation}
where $\big( L^2 (\mathbb{R}^0) \big)_s = \mathbb{C}$. Here, $s$ denotes the subspace of symmetric functions. A vector $\psi$ in $\F_X$ is a sequence $\psi = \{ \psi^{(n)} \}_{n \geq 0}$ of $n$-particle wavefunctions $\psi^{(n)} \in \big( L^2 (\mathbb{R}^{3n}) \big)_s$. The scalar product between $\psi_a, \psi_b \in \F_X$ is defined by
\begin{equation}
\langle \psi_a, \psi_b \rangle_{\F_X} = \sum_{n \geq 0} \langle \psi_a^{(n)}, \psi_b^{(n)} \rangle_{L^2 (\mathbb{R}^{3n})}
\end{equation}
and we will omit the subscript $\F_X$ from now on. We let
\begin{equation}
\Omega := \{1, 0, 0, \cdots \} \in \F_X,
\end{equation}
which is called the vacuum. We will also make use of the space
\begin{equation}
\F := L^{\infty}([0, T], \F_X).
\end{equation}

For $n \geq 1$, let
\begin{equation}
F_n := \{ \xi_n (t, X_n) \; | \; \xi_n \in L_t^{\infty} (L_{X_n}^2)_s \cap \bigcap_{i<j}^n L_t^2 L_{x_i - x_j}^6 L_{x_i + x_j}^2 L_{X_n \backslash \{x_i, x_j \} }^2 \}
\end{equation}
with
\begin{equation}
\| \xi_n \|_{F_n} = \| \xi_n (t, X_n) \|_{L_t^{\infty} L_{X_n}^2} + \max_{1 \leq i < j \leq n} \{ \| \xi_n (t, X_n) \|_{L_t^2 L_{x_i - x_j}^6 L_{x_i + x_j}^2 L_{X_n \backslash \{x_i, x_j \} }^2} \},
\end{equation}
and let $F_0 = \mathbb{C}$. Here, we again consider a fixed time interval $t \in [0, T]$ only. We define another space
\begin{equation}
\widetilde{\F} := \bigoplus_{n \geq 0} F_n,
\end{equation}
equipped with the norm
\begin{equation}
\| \cdot \|_{\widetilde{\F}}^2 = \sum_{n \geq 0} \| \cdot \|_{F_n}^2.
\end{equation}

On $\F_X$, the creation operator $a_x^{\dag}$ and the annihilation operator $a_x$ for $x \in \mathbb{R}^3$ are defined by
\begin{align}
(a_x^{\dag} \psi)^{(n)}(x_1, \cdots, x_n ) &= \frac{1}{\sqrt n} \sum_{j=1}^n \delta(x-x_j) \psi^{(n-1)}(x_1, \cdots, x_{j-1}, x_{j+1}, \cdots, x_n), \\
(a_x \psi)^{(n)}(x_1, \cdots, x_n ) &= \sqrt{n+1} \; \psi^{(n+1)}(x, x_1, \cdots, x_n).
\end{align}
For $f \in L^2(\mathbb{R}^3)$, $a^{\dag}(f)$ and $a(f)$ are given by
\begin{align}
a^{\dag}(f) &= \int dx f(x) a_x^{\dag}, \\
a(f) &= \int dx \overline{f(x)} a_x.
\end{align}
The creation operator $a^{\dag}(f)$ is the adjoint of the annihilation operator $a(f)$, and they satisfy the canonical commutation relations
\begin{equation}
[a(f), a^{\dag}(g)] = \langle f, g \rangle_{L^2(\mathbb{R}^3)}, \quad [a(f), a(g)] = [a^{\dag} (f), a^{\dag} (g)] = 0.
\end{equation}

The number operator $\N$ on $\F_X$ is defined by
\begin{equation}
\N := \int dx \: a_x^{\dag} a_x,
\end{equation}
and it also satisfies
\begin{equation}
(\N \psi)^{(n)} = n \psi^{(n)}.
\end{equation}
We define the Hamiltonian $\h_N$ on $\F_X$ by $(\h_N \psi)^{(n)} = H_n \psi^{(n)}$. Using the operators $a_x$ and $a_x^{\dag}$, it can be rewritten as
\begin{equation}
\h_N = \int dx \; a_x^{\dag} (-\Delta_x ) a_x + \frac{1}{2N} \iint dx dy \; V(x-y) a_x^{\dag} a_y^{\dag} a_y a_x.
\end{equation}

For $f \in L^2(\mathbb{R}^3)$, the Weyl operator $W(f)$ and the coherent state $\psi(f)$ are given by
\begin{equation}
W(f) := \textrm{exp}(a^{\dag}(f) - a(f)),
\end{equation}
and
\begin{equation}
\psi(f) := W(f) \Omega.
\end{equation}
The Weyl operator $W(f)$ satisfies
\begin{equation}
W(f) = e^{-\|f\|^2 /2} \textrm{exp}(a^{\dag}(f)) \textrm{exp}(-a(f)).
\end{equation}
Important properties of Weyl operators are collected in Lemma 2.2 of \cite{RS}.

\subsection{Main theorem}

In this paper, for factorized initial data and the interaction potentials $V \in L^3 (\mathbb{R}^3 )+ L^{\infty} (\mathbb{R}^3 )$, we extend the convergence rate $O(1/N^{\frac{1}{2}})$ in Rodnianski and Schlein \cite{RS} to $O(1/N)$. More precisely, we prove the following theorem.

\begin{thm} \label{main theorem}
Suppose that $V = V_1 + V_2$ where $V_1 \in L^3 (\mathbb{R}^3)$ and $V_2 \in L^{\infty} (\mathbb{R}^3 )$. Let $\gamma_{N, t}^{(1)}$ be the one-particle marginal density associated with the time evolution of the factorized initial state $\{0, \cdots, 0, \varphi^{\otimes N}, 0, \cdots \}$ as in \eqref{marginal density}. Then there exist constants $C$ and $K$, depending only on $\| \varphi \|_{H^1}$, $\| V_1 \|_{L^3}$, and $\| V_2 \|_{L^{\infty}}$ such that
\begin{equation}
\textrm{Tr } \Big| \gamma_{N, t}^{(1)} - | \varphi_t \rangle \langle \varphi_t | \Big| \leq \frac{C e^{Kt}}{N}. \label{trace norm bound}
\end{equation}
\end{thm}

\begin{remark}
Using the same arguments used to prove \eqref{trace norm bound}, we can extend the result to the higher marginals $\gamma_{N, t}^{(k)}$.
\end{remark}

\begin{remark} \label{time interval}
For the proof of Theorem \ref{main theorem}, we first assume that $t \in [0, T]$ and prove the bound
\begin{equation}
\textrm{Tr } \Big| \gamma_{N, t}^{(1)} - | \varphi_t \rangle \langle \varphi_t | \Big| \leq \frac{C e^{KT}}{N}.
\end{equation}
Letting $T = t$, Theorem \ref{main theorem} follows.
\end{remark}

\begin{remark} \label{HS norm}
What we actually prove in this paper is
\begin{equation}
\Big| \gamma_{N, t}^{(1)} - | \varphi_t \rangle \langle \varphi_t | \Big|_{HS} \leq \frac{C e^{Kt}}{N}, \label{HS norm bound}
\end{equation}
which implies \eqref{trace norm bound}. Here, $\| \cdot \|_{HS}$ denotes the Hilbert-Schmidt norm. The argument for this implication can be found in Remark 1.4 of \cite{RS}.
\end{remark}

\begin{remark}
Note that the nonlinear Hartree equation \eqref{Hartree} is globally well-posed when $V$ satisfies the assumptions of Theorem \ref{main theorem}. See Remark 1.3 of \cite{RS} for further detail.
\end{remark}

In \cite{RS}, the authors had proved the convergence rate is $1/N$ with initial data is coherent state. Then, by writing the factorized state
\begin{equation} 
\{0, \cdots, 0, \varphi^{\otimes N}, 0, \cdots \}= \frac{(a^{\dag}(\varphi))^N}{\sqrt{N!}} \Omega = d_N \int_0^{2\pi} \frac{d\theta}{2\pi} e^{i\theta N} W(e^{-i\theta\sqrt{N}\varphi}) \Omega
\end{equation}
with the constant
\begin{equation}
d_N := \dfrac{\sqrt{N!}}{N^{N/2}e^{-N/2}} \simeq N^{1/4},
\end{equation}
they could prove that the rate of convergence is of order $1/\sqrt{N}$.

To explain our idea more explicitly, we need first to give a brief outline of the proof of the main theorem in \cite{RS}.

\subsection{Outline of the idea in \cite{RS}}

Let $\Gamma_{N, t}^{(1)}(x, y)$ be the kernel of the one particle marginal density associated with the time evolution of the coherent state $W(\sqrt{N}\varphi) \Omega$, which is defined by
\begin{equation}
\Gamma_{N, t}^{(1)} (x, y) = \frac{1}{N} \langle e^{-i \h_N t}W(\sqrt{N} \varphi) \Omega, a_y^{\dag} a_x e^{-i \h_N t} W(\sqrt N \varphi)\Omega\rangle,
\end{equation}
We expect that the limit of the kernel of one particle marginal density is $\overline \varphi_t(x) {\varphi_t}(y)$, thus we expand $\Gamma_{N, t}^{(1)}(x, y)$ in terms of $(a_x - \sqrt N \varphi_t (x))$ and $(a_y^{\dag} - \sqrt N \overline{\varphi_t} (y))$. Then, we get
\begin{align}
\Gamma_{N, t}^{(1)}(x, y) &= \varphi_t (x) \overline{\varphi_t} (y) \nonumber \\
& \quad + \frac{1}{N} \langle \Omega, W^{\dag}(\sqrt{N} \varphi) e^{i \h_N t} (a_y^{\dag} - \sqrt N \overline{\varphi_t} (y)) (a_x - \sqrt N \varphi_t (x)) e^{-i \h_N t} W(\sqrt N \varphi)\Omega\rangle \nonumber \\
& \quad + \frac{\varphi_t (x)}{\sqrt N} \langle \Omega, W^{\dag}(\sqrt{N} \varphi) e^{i \h_N t} (a_y^{\dag} - \sqrt N \overline{\varphi_t} (y) ) e^{-i \h_N t} W(\sqrt N \varphi) \Omega \rangle \label{one particle marginal 1} \\
& \quad + \frac{\overline{\varphi_t} (y)}{\sqrt N} \langle \Omega, W^{\dag}(\sqrt{N} \varphi) e^{i \h_N t} (a_x - \sqrt N \varphi_t (x) ) e^{-i \h_N t} W(\sqrt N \varphi)\Omega\rangle. \nonumber
\end{align}

It was shown by Hepp \cite{H} (also by Ginibre and Velo \cite{GV1, GV2}) that
\begin{equation}
W^{\dag} (\sqrt N \varphi_s ) e^{i \h_N (t-s)} (a_x - \sqrt N \varphi_t (x) ) e^{-i \h_N (t-s)} W(\sqrt N \varphi_s) = \U^{\dag}(t;s) a_x \U(t;s),
\end{equation}
where $\U(t;s)$ is a unitary operator defined through
\begin{equation}
\U(t;s) := W^{\dag} (\sqrt N \varphi_t ) e^{-i \h_N (t-s)} W (\sqrt N \varphi_s ).
\end{equation}
The operator $\U (t;s)$ also satisfies
\begin{equation}
i \partial_t \U(t;s) = (H_2 (t) + H_3 (t) + H_4 ) \U(t;s), \qquad \U(s;s) = I, \label{evolution of U}
\end{equation}
where the generators $H_2$, $H_3$, and $H_4$ are defined as follows:
\begin{align}
H_2 (t) &:= \int dx \nabla_x a_x^{\dag} \nabla_x a_x + \int dx (V*|\varphi_t |^2 )(x) a_x^{\dag} a_x + \iint dx dy V(x-y) \overline{\varphi_t} (x) \varphi_t (y) a_y^{\dag} a_x \nonumber \\
& \quad + \frac{1}{2} \iint dx dy V(x-y) ( \varphi_t (x) \varphi_t (y) a_x^{\dag} a_y^{\dag} + \overline{\varphi_t} (x) \overline{\varphi_t} (y) a_x a_y ),  \label{H2} \\
H_3 (t) &:= \frac{1}{\sqrt N} \iint dx dy V(x-y) \varphi_t (y) a_x^{\dag} a_y^{\dag} a_x + \frac{1}{\sqrt N} \iint dx dy V(x-y) \overline{\varphi_t} (y)
 a_x^{\dag} a_y a_x, \label{H3} \\
H_4 &:= \frac{1}{2N} \iint dx dy V(x-y) a_x^{\dag} a_y^{\dag} a_x a_y. \label{H4}
\end{align}
Using $\U(t;s)$, we can rewrite \eqref{one particle marginal 1} as
\begin{align}
& \Gamma_{N, t}^{(1)}(x, y) - \varphi_t (x) \overline{\varphi_t} (y) \label{one particle marginal 2} \\
&= \frac{1}{N} \langle \Omega, \U^{\dag}(t;0) a_y^{\dag} a_x \U(t;0) \Omega \rangle \nonumber \\
& \quad + \frac{\varphi_t (x)}{\sqrt N} \langle \Omega, \U^{\dag}(t;0) a_y^{\dag} \U(t;0) \Omega \rangle + \frac{\overline{\varphi_t} (y)} {\sqrt N} \langle \Omega, \U^{\dag}(t;0) a_x \U(t;0) \Omega \rangle. \nonumber
\end{align}

Proposition 4.1 of \cite{GV1} shows that there exists a unique group of unitary operators $\widetilde{\U}_2 (t;s)$ satisfying
\begin{equation}
i \partial_t \widetilde{\U}_2 (t;s) = (\widetilde{H}_2 (t) - H_0 ) \widetilde{\U}_2 (t;s), \qquad \widetilde{\U}_2 (s;s) = I,
\end{equation}
where $\widetilde{H}_2 (t)$ is defined by
\begin{equation}
\widetilde{H}_2 (t) := e^{i H_0 t} H_2 (t) e^{-i H_0 t}.
\end{equation}
We define
\begin{equation}
\U_2 (t;s) := e^{-i H_0 t} \widetilde{\U}_2 (t;s) e^{i H_0 s}.
\end{equation}
We note that $\U_2$ is well-defined on $\F_X$, since the free evolutions $e^{-i H_0 t}$ and $e^{i H_0 s}$ are well-defined on $\F_X$, and the unitary operator $\widetilde{\U}_2 (t;s)$ is also well-defined on $\F_X$. Since the operators $\widetilde{\U}_2$, $e^{-i H_0 t}$, and $e^{i H_0 s}$ are bounded in $\D( \N^{\delta} )$ for any $\delta \in \mathbb{R}$, so is $\U_2$. 

Furthermore, it can be proved from Proposition 2.2 of \cite{GV2} and from Lemma \ref{stability} that $\U_2$ is strongly differentiable from $\Q (H_0 + \N)$, the form domain of $H_0 + \N$, to its dual $\Q^* (H_0 + \N)$. Thus, on a dense subset $\D( H_0 ) \cap \D (\N)$ of $\F_X$, we can check from the definition that
\begin{equation}
i \partial_t \U_2 (t;s) = H_2 (t) \U_2 (t;s), \qquad \U_2 (s;s) = I \label{evolution of U_2}.
\end{equation}
For simplicity, we will use notations
\begin{equation}
\U(t) := \U(t;0), \;\;\; \U_2 (t) := \U_2 (t;0).
\end{equation}

To estimate the difference between $\Gamma_{N, t}^{(1)}(x, y)$ and $\varphi_t (x) \overline{\varphi_t} (y)$ as in \eqref{one particle marginal 2}, one needs to find a good approximation of $\U(t;0)$, which conserves the parity of the number of particles, and we call it $\widetilde{\U}(t;0)$. Then
\begin{align}
& \Gamma_{N, t}^{(1)}(x, y) - \varphi_t (x) \overline{\varphi_t} (y) \nonumber \\
&= \frac{1}{N} \langle \Omega, \U^{\dag}(t;0) a_y^{\dag} a_x \U(t;0) \Omega \rangle \\
& \quad + \frac{\varphi_t (x)}{\sqrt N} \Big( \langle \Omega, \U^{\dag}(t;0) a_y^{\dag} \big( \U(t;0) - \widetilde{\U}(t;0) \big) \Omega \rangle + \langle \Omega, \big( \U^{\dag}(t;0) - \widetilde{\U}^{\dag}(t;0)\big) a_y^{\dag} \widetilde{\U}(t;0) \Omega \rangle \Big) \nonumber \\
& \quad + \frac{\overline{\varphi_t} (y)} {\sqrt N} \Big( \langle \Omega, \U^{\dag}(t;0) a_x \big( \U(t;0) - \widetilde{\U}(t;0) \big) \Omega \rangle + \langle \Omega, \big( \U^{\dag}(t;0) - \widetilde{\U}^{\dag}(t;0)\big) a_x \widetilde{\U}(t;0) \Omega \rangle \Big). \nonumber
\end{align}
If we define $\widetilde{\U}$ through
\begin{equation}
i \partial_t \widetilde{\U}(t;0) = (H_2+H_4) \widetilde{\U}(t;0), \qquad \widetilde{\U}(0;0) = I, \label{evolution of tilde U in introduction}
\end{equation}
then it can be proved that
\begin{equation}
\| \U(t;0)\Omega - \widetilde{\U}(t;0) \Omega \| \leq \frac{C e^{Kt}}{\sqrt{N}}
\end{equation}
and
\begin{equation}
\langle \U(t;0) \Omega, \N \U(t;0) \Omega \rangle \leq C e^{Kt},
\end{equation}
which are Lemma 3.9 and Proposition 3.3 of \cite{RS}, respectively. Thus,
\begin{equation}
\textrm{Tr} \Big| \: \Gamma_{N, t}^{(1)} - | \varphi_t \rangle \langle \varphi_t | \Big| \leq \frac{C e^{Kt}}{N}.
\end{equation}
Note that the results in \cite{RS} are under the assumption that $|V(x)|^2 \leq D(1 - \Delta_x)$ for some constant $D$. In this paper, we will use these estimates, since the potential $V$ in Theorem \ref{main theorem} satisfies the same assumption.

\subsection{Factorized initial data}

Recall that
\begin{equation}
W^{\dag} (\sqrt N \varphi ) e^{i \h_N t} (a_x - \sqrt N \varphi_t (x) ) e^{-i \h_N t} W(\sqrt N \varphi) = \U^{\dag}(t) a_x \U(t). \label{Weyl operator identity}
\end{equation}
By definition, we have that
\begin{equation}
\Big( W(\sqrt{N} \varphi) \Omega \Big)^{(N)} = e^{-N/2} \frac{\big(a^{\dag} (\sqrt{N} \varphi) \big)^N}{N!} \Omega = \frac{1}{d_N} \frac{\big(a^{\dag} (\varphi) \big)^N}{\sqrt{N!}} \Omega. \label{factorized data}
\end{equation}

For factorized initial data, it follows from \eqref{Weyl operator identity} and \eqref{factorized data} that
\begin{align}
& \gamma^{(1)}_{N,t}(x;y) = \frac{1}{N} \left \langle \frac{(a^{\dag}(\varphi))^N}{\sqrt{N!}} \Omega, e^{i\h_N t} a_y^{\dag} a_x e^{-i\h_N t} \frac{(a^{\dag}(\varphi))^N}{\sqrt{N!}} \Omega \right \rangle \nonumber \\
&= \frac{d_N^2}{N} \left \langle e^{i\h_N t} a_y e^{-i\h_N t} \big( W(\sqrt{N} \varphi) \Omega \big)^{(N)}, e^{i\h_N t} a_x e^{-i\h_N t} \big( W(\sqrt{N} \varphi) \Omega \big)^{(N)} \right \rangle \nonumber \\
&= \frac{d_N^2}{N} \left \langle \left( e^{i\h_N t} a_y e^{-i\h_N t} W(\sqrt{N} \varphi) \Omega \right)^{(N-1)}, \left( e^{i\h_N t} a_x e^{-i\h_N t} W(\sqrt{N} \varphi) \Omega \right)^{(N-1)} \right \rangle \\
&= \frac{d_N^2}{N} \Bigg \langle \Big( W(\sqrt{N} \varphi) \U^{\dag}(t) \big( a_y + \sqrt N \varphi_t (y) \big) \U(t) \Omega \Big)^{(N-1)}, \nonumber \\
& \qquad \qquad \Big( W(\sqrt{N} \varphi) \U^{\dag}(t) \big( a_x + \sqrt N \varphi_t (x) \big) \U(t) \Omega \Big)^{(N-1)} \Bigg \rangle. \nonumber
\end{align}
Thus, we obtain the following equation for one-particle marginal.
\begin{align}
& \quad \gamma^{(1)}_{N,t}(x;y) - \overline\varphi_t (y) \varphi_t(x) \nonumber \\
&= \frac{d_N^2}{N} \left \langle \Big( W(\sqrt{N} \varphi) \U^{\dag}(t) a_y \U(t) \Omega \Big)^{(N-1)}, \Big( W(\sqrt{N} \varphi) \U^{\dag}(t) a_x \U(t) \Omega \Big)^{(N-1)} \right\rangle \label{expansionU} \\
& \quad + \overline\varphi_t (y) \frac{d_N^2}{\sqrt{N}} \left \langle \Big( W(\sqrt{N} \varphi) \Omega \Big)^{(N-1)},
\Big( W(\sqrt{N} \varphi) \U^{\dag}(t) a_x \U(t) \Omega \Big)^{(N-1)} \right \rangle \nonumber \\
& \quad + \varphi_t(x) \frac{d_N^2}{\sqrt{N}} \left \langle \Big( W(\sqrt{N} \varphi) \U^{\dag}(t) a_y \U(t) \Omega \Big)^{(N-1)}, \Big( W(\sqrt{N} \varphi) \Omega \Big)^{(N-1)} \right \rangle. \nonumber
\end{align}

Define
\begin{equation}
E(t, x, y) := \gamma^{(1)}_{N,t}(x;y) - \overline\varphi_t (y) \varphi_t(x).
\end{equation}
We also define $E_2 (t, x, y)$ by putting $\U_2$ instead of $\U$ in \eqref{expansionU} as follows.
\begin{align}
E_2 (t, x, y) &= \frac{d_N^2}{N} \left \langle \Big( W(\sqrt{N} \varphi) \U_2^{\dag}(t) a_y \U_2(t) \Omega \Big)^{(N-1)}, \Big( W(\sqrt{N} \varphi) \U_2^{\dag}(t) a_x \U_2(t) \Omega \Big)^{(N-1)} \right \rangle \nonumber \\
& \quad + \overline\varphi_t (y) \frac{d_N^2}{\sqrt{N}} \left \langle \Big( W(\sqrt{N} \varphi) \Omega \Big)^{(N-1)},
\Big( W(\sqrt{N} \varphi) \U_2^{\dag}(t) a_x \U_2(t) \Omega \Big)^{(N-1)} \right \rangle \label{expansion U_2} \\
& \quad + \varphi_t(x) \frac{d_N^2}{\sqrt{N}} \left \langle \Big( W(\sqrt{N} \varphi) \U_2^{\dag}(t) a_y \U_2(t) \Omega \Big)^{(N-1)}, \Big( W(\sqrt{N} \varphi) \Omega \Big)^{(N-1)} \right \rangle. \nonumber
\end{align}

\subsection{Arrangement of the paper}

This paper is organized as follows:

\begin{enumerate}

\item
We will first study the evolution of $\U_2^{\dag}(t) a_x \U_2(t)$ in section 2, then, by using it, we will prove the following proposition.

\begin{prop} \label{E_2 bound}
Let $E_2 (t, x, y)$ be defined by \eqref{expansion U_2}. Then, there exist constants $C$ and $K$ such that
\begin{equation}
\| E_2 (t, x, y) \|_{L_{x, y}^2} \leq \frac{Ce^{Kt}}{N}.
\end{equation}
\end{prop}

\item
To estimate the difference $\U(t)-\U_2(t)$, we study more regularity for $\U_2(t)$ dynamics in
section 3. More precisely, we will prove the following proposition.

\begin{prop} \label{L6 norm estimate}
Let $\psi_2(t) := \U_2(t;s) \psi$ be the unique solution of initial value problem
\begin{equation}
	\begin{cases}
		i \partial_t (\N^j \psi_2(t) ) = \N^j H_2(t) \psi_2(t) \\
		\psi_2(s) = \psi \in \F_X
	\end{cases}. \label{definition of psi_2}
\end{equation}
Then there exists a constant $C$ depending only on $j$, $\| V_1 \|_{L^3}$, $\| V_2 \|_{L^{\infty}}$ and $\| \varphi \|_{H^1}$, such that
\begin{equation}
\| \N^j \psi_2 \|_{\widetilde{\F}} \leq C \|\N^j \psi\| + C T \| (\N+2)^{j+1} \psi_2 \|_{\F}.
\end{equation}
\end{prop}

\item
Let
\begin{equation}
\R_y (t) := \U^{\dag}(t) a_y \U(t) - \U_2^{\dag}(t) a_y \U_2(t). \label{definition of R_y}
\end{equation}
In Section 4, using Proposition \ref{L6 norm estimate}, we prove the following proposition, which gives an estimate for the difference between $\U(t)$ dynamics and $\U_2(t)$ dynamics.

\begin{prop} \label{R_y estimate}
Let $\R_y(t)$ be defined as in \eqref{definition of R_y}. Then, for all $j \in \mathbb{N}$, there exist constants $C$ and $K$ such that
\begin{equation}
\left( \int dy \| (\N+1)^{\frac{j}{2}} \R_y (t) \Omega \|^2 \right)^{\frac{1}{2}} \leq \frac{C e^{KT}}{\sqrt N}.
\end{equation}
\end{prop}

\item
In Section 5, we prove the main theorem, using Proposition \ref{E_2 bound}, Proposition \ref{L6 norm estimate}, and Proposition \ref{R_y estimate}. Some technical estimates are proved in Section 6.

\end{enumerate}

\section{Evolution of $\U_2^{\dag}(t) a_x \U_2(t)$}

\subsection{Properties of $\U_2^{\dag}(t) a_x \U_2(t)$}
We first want to obtain some algebraic properties of an operator $\U_2^{\dag}(t) a_x \U_2(t)$, which will act on the vacuum $\Omega$. For simplicity, we let
\begin{equation}
a_x (t) := \U_2^{\dag}(t) a_x \U_2(t),
\end{equation}
\begin{equation}
a_x^{\dag} (t) := \U_2^{\dag}(t) a_x^{\dag} \U_2(t).
\end{equation}
Formal calculation shows that the time derivatives of $a_x(t)$ and $a_x^{\dag}(t)$ are given by
\begin{equation}
i \partial_t
	\left(
	\begin{array}{c}
		a_x(t) \\
		a_x^{\dag}(t) \\
	\end{array}
	\right)
= \U_2^{\dag}(t) \Big[
	\left(
	\begin{array}{c}
		a_x \\
		a_x^{\dag} \\
	\end{array}
	\right)
, H_2 \Big] \U_2(t). \label{derivative of a_x}
\end{equation}
Commutators in the right hand side of \eqref{derivative of a_x} can be calculated explicitly as follows.
\begin{equation*}
\Big[ a_x , \int dy \nabla_y a_y^{\dag} \nabla_y a_y \Big] = -\Delta_x a_x,
\end{equation*}
\begin{equation*}
\Big[ a_x^{\dag} , \int dy \nabla_y a_y^{\dag} \nabla_y a_y \Big] = \Delta_x a_x^{\dag},
\end{equation*}
\begin{equation*}
\Big[ a_x , \int dy (V*|\varphi_t|^2)(y) a_y^{\dag} a_y \Big] = (V*|\varphi_t|^2)(x) a_x,
\end{equation*}
\begin{equation*}
\Big[ a_x^{\dag} , \int dy (V*|\varphi_t|^2)(y) a_y^{\dag} a_y \Big] = -(V*|\varphi_t|^2)(x) a_x^{\dag},
\end{equation*}
\begin{equation*}
\Big[ a_x , \iint dy dz V(y-z) \overline{\varphi_t}(y) \varphi_t(z) a_z^{\dag} a_y \Big] = \int dy V(x-y) \overline{\varphi_t}(y) \varphi_t (x) a_y,
\end{equation*}
\begin{equation*}
\Big[ a_x^{\dag} , \iint dy dz V(y-z) \overline{\varphi_t}(y) \varphi_t(z) a_z^{\dag} a_y \Big] = -\int dy V(x-y)
\overline{\varphi_t}(x) \varphi_t (y) a_y^{\dag},
\end{equation*}
\begin{equation*}
\Big[ a_x , \frac{1}{2} \iint dy dz V(y-z) \varphi_t(y) \varphi_t(z) a_y^{\dag} a_z^{\dag} \Big] = \int dy V(x-y)
\varphi_t(x) \varphi_t (y) a_y^{\dag},
\end{equation*}
\begin{equation*}
\Big[ a_x^{\dag} , \frac{1}{2} \iint dy dz V(y-z) \overline{\varphi_t}(y) \overline{\varphi_t}(z) a_y a_z \Big] = -\int dy V(x-y) \overline{\varphi_t}(x) \overline{\varphi_t}(y) a_y.
\end{equation*}

Now, \eqref{derivative of a_x} becomes
\begin{align}
& \;\;\;\; i \partial_t
	\left(
  	\begin{array}{c}
    	a_x (t)\\
        a_x^{\dag} (t) \\
    \end{array}
	\right) \nonumber \\
&= \U_2^{\dag}(t)
	\left(
	\begin{array}{c}
  		-\Delta_x a_x + (V*|\varphi_t|^2)(x) a_x  \\
  		\Delta_x a_x^{\dag} - (V*|\varphi_t|^2)(x) a_x^{\dag} \\
	\end{array}
	\right)
\U_2(t) \nonumber \\
& \quad + \U_2^{\dag}(t)
	\left(
	\begin{array}{c}
		\int dy V(x-y) \overline{\varphi_t}(y) \varphi_t (x) a_y + \int dy V(x-y) \varphi_t(x) \varphi_t (y) a_y^{\dag} \\
		- \int dy V(x-y) \overline{\varphi_t}(x) \varphi_t (y) a_y^{\dag} - \int dy V(x-y) \overline{\varphi_t}(x) \overline{\varphi_t}(y) a_y \label{differential equation for a_x} \\
	\end{array}
	\right)
\U_2(t) \\
&=  \left(
	\begin{array}{cc}
		-\Delta_x + (V*|\varphi_t|^2)(x) & 0 \\
		0 & \Delta_x - (V*|\varphi_t|^2)(x) \\
	\end{array}
	\right)
	\left(
	\begin{array}{c}
		a_x (t) \\
		a_x^{\dag} (t) \\
	\end{array}
	\right) \nonumber \\
& \quad + \int dy
	\left(
	\begin{array}{cc}
		V(x-y) \overline{\varphi_t}(y) \varphi_t (x) & V(x-y) \varphi_t(x) \varphi_t (y) \\
		-V(x-y) \overline{\varphi_t}(x) \overline{\varphi_t}(y) & - V(x-y) \overline{\varphi_t}(x) \varphi_t (y) \\
	\end{array}
	\right)
	\left(
	\begin{array}{c}
		a_y (t) \\
		a_y^{\dag} (t) \\
	\end{array}
	\right). \nonumber
\end{align}

Using \eqref{differential equation for a_x}, we can immediately see that
\begin{equation}
	\left(
	\begin{array}{c}
		a_x (t) \\
		a_x^{\dag} (t) \\
	\end{array}
	\right)
= G(t)
	\left(
	\begin{array}{c}
		a_x \\
		a_x^{\dag} \\
	\end{array}
	\right)
:=	\left(
	\begin{array}{cc}
		G_1 & G_2 \\
		G_3 & G_4 \\
	\end{array}
	\right)
	\left(
	\begin{array}{c}
		a_x \\
		a_x^{\dag} \\
	\end{array}
	\right).
\end{equation}
Here, $G_1$, $G_2$, $G_3$, and $G_4$ depend on $t$, but we omitted it. Since
\begin{equation}
(\U_2^{\dag}(t) a_x \U_2(t))^{\dag} = \U_2^{\dag}(t) a_x^{\dag} \U_2(t),
\end{equation}
we can easily see that $G_4 = \overline{G_1}$ and $G_3 = \overline{G_2}$. Thus,
\begin{equation}
	\left(
	\begin{array}{c}
		a_x (t) \\
		a_x^{\dag} (t) \\
	\end{array}
	\right)
=	\left(
	\begin{array}{cc}
		G_1 & G_2 \\
		\overline{G_2} & \overline{G_1} \\
	\end{array}
	\right)
	\left(
	\begin{array}{c}
		a_x \\
		a_x^{\dag} \\
	\end{array}
	\right). \label{evolution of a_x}
\end{equation}
From \eqref{evolution of a_x}, we can obtain the following lemma, which shows an important property of the operator $\U_2^{\dag}(t) a_x \U_2(t)$.

\begin{lem} \label{G_2 bound}
Let
\begin{equation}
G_2 (t, x) := \U_2^{\dag}(t) a_x \U_2(t) \Omega.
\end{equation}
Then, $G_2 (t, x) \in L^2 (\mathbb{R}^3, L^2 (\mathbb{R}^3))$ for any fixed $t > 0$. Furthermore, if we let 
\begin{equation} \label{definition of G_2}
G_2 (t, x, z) := (\U_2^{\dag}(t) a_x \U_2(t) \Omega) (z),
\end{equation}
then we have the following bound for $G_2$:
\begin{equation}
\| G_2 (t, x, z) \|_{L_{x, z}^2} \leq C e^{Kt}.
\end{equation}
\end{lem}

\textit{Proof.}
We first observe that
\begin{equation} \label{a_x Omega bound 1}
\int dx \langle a_x(t) \Omega, a_x(t) \Omega \rangle = \int dx \langle G_2 (t, x), G_2 (t, x) \rangle.
\end{equation}
On the other hand, we have from the definition and from Lemma \ref{U_2 conserves N} that
\begin{equation} \label{a_x Omega bound 2}
\int dx \langle a_x(t) \Omega, a_x(t) \Omega \rangle = \int dx \langle a_x \U_2(t) \Omega, a_x \U_2(t) \Omega \rangle = \langle \U_2(t) \Omega, \N \U_2(t) \Omega \rangle \leq Ce^{Kt}.
\end{equation}
Thus, from \eqref{a_x Omega bound 1} and \eqref{a_x Omega bound 2}, we can find that $G_2 (t, x) \in L^2 (\mathbb{R}^3, F_X)$ for any fixed $t > 0$. In particular, $\U_2^{\dag}(t) a_x \U_2(t) \Omega$ is well-defined for almost every $x$. Let $\eta^{(n)}$ be a function in $L^2 (\mathbb{R}^{3n})$ with $n \geq 0$. If $n \neq 1$, then the formal calculation \eqref{evolution of a_x} shows that
\begin{equation}
\langle \eta^{(n)}, \U_2^{\dag}(t) a_x \U_2(t) \Omega \rangle = 0.
\end{equation}
Thus, we can see that $G_2 (t, x) \in L^2 (\mathbb{R}^3, L^2 (\mathbb{R}^3))$, which proves the first part of the lemma. Now, we rewrite \eqref{a_x Omega bound 1} and \eqref{a_x Omega bound 2} as
\begin{equation}
\int dx \langle a_x(t) \Omega, a_x(t) \Omega \rangle = \iint dx dz |G_2 (t, x, z)|^2 \leq Ce^{Kt}.
\end{equation}
This completes the proof of the lemma.

\subsection{Proof of Proposition \ref{E_2 bound}.} 

Using the properties of $a_x (t)$, we can prove Proposition \ref{E_2 bound}.

\textit{Proof of Proposition \ref{E_2 bound}.}
Recall that
\begin{align}
& \quad E_2 (t, x, y) \nonumber \\
&= \frac{d_N^2}{N} \left \langle \Big( W(\sqrt{N} \varphi) \U_2^{\dag}(t) a_y \U_2(t) \Omega \Big)^{(N-1)}, \Big( W(\sqrt{N} \varphi) \U_2^{\dag}(t) a_x \U_2(t) \Omega \Big)^{(N-1)} \right\rangle \label{recall expansion U_2} \\
& \quad + \overline\varphi_t (y) \frac{d_N^2}{\sqrt{N}} \left \langle \Big( W(\sqrt{N} \varphi) \Omega \Big)^{(N-1)},
\Big( W(\sqrt{N} \varphi) \U_2^{\dag}(t) a_x \U_2(t) \Omega \Big)^{(N-1)} \right \rangle \nonumber \\
& \quad + \varphi_t(x) \frac{d_N^2}{\sqrt{N}} \left \langle \Big( W(\sqrt{N} \varphi) \U_2^{\dag}(t) a_y \U_2(t) \Omega \Big)^{(N-1)}, \Big( W(\sqrt{N} \varphi) \Omega \Big)^{(N-1)} \right \rangle. \nonumber
\end{align}

We first want to compute $\Big( W(\sqrt{N} \varphi) \U_2^{\dag}(t) a_y \U_2(t) \Omega \Big)^{(N-1)}$ explicitly. We have already seen from Lemma \ref{G_2 bound} that $a_x (t) \Omega \in L^2 (\mathbb{R}^3, L^2 (\mathbb{R}^3) )$ and we write it as
\begin{equation}
a_x (t) \Omega = \U_2^{\dag} (t) a_x \U_2(t) \Omega = G_2 (t, x).
\end{equation}
Now, we calculate from the definition of the Weyl operator that
\begin{align}
& \Big( W(\sqrt N \varphi) \U_2^{\dag} (t) a_x \U_2(t) \Omega \Big)^{(N-1)} \nonumber \\
&= e^{-\frac{N}{2}} \Big( \exp \big( a^{\dag} (\sqrt N \varphi ) \big) \exp \big( -a(\sqrt N \varphi ) \big) G_2 (t, x) \Big)^{(N-1)} \nonumber \\
&= e^{-\frac{N}{2}} \Big( \exp \big( a^{\dag} (\sqrt N \varphi ) \big) \big( G_2 (t, x) - \sqrt N \int dz \; G_2 (t, x, z) \overline{\varphi}(z) \big) \Big)^{(N-1)} \label{U_2 Omega} \\
&= e^{-\frac{N}{2}} \Big( \frac{1}{(N-2)!} \big(a^{\dag} (\sqrt N \varphi ) \big)^{N-2} G_2 (t, x) \nonumber \\
& \qquad \qquad - \frac{\sqrt N}{(N-1)!} \int dz \; G_2 (t, x, z) \overline{\varphi}(z) \big(a^{\dag} (\sqrt N \varphi ) \big)^{N-1} \Omega \Big), \nonumber
\end{align}
where we used the notation $G_2 (t, x, z)$, which was defined in \eqref{definition of G_2}. Thus,
\begin{align}
& \Big( W(\sqrt N \varphi) \U_2^{\dag} (t) a_x \U_2(t) \Omega \Big)^{(N-1)} (x_1, x_2, \cdots, x_{N-1}) \\
&= \frac{1}{d_N \sqrt N} \Big( \sum_{i=1}^{N-1} G_2(t, x, x_i) \prod_{j \neq i}^{N-1} \varphi(x_j) - N \int dz \; G_2 (t, x, z) \overline{\varphi}(z) \prod_{j=1}^{N-1} \varphi(x_j) \Big). \nonumber
\end{align}

We also know that
\begin{equation}
\Big( W(\sqrt N \varphi) \Omega \Big)^{(N-1)}(x_1, \cdots, x_{N-1}) = \frac{1}{d_N} \prod_{j=1}^{N-1} \varphi(x_j). \label{W Omega}
\end{equation}
Using \eqref{U_2 Omega} and \eqref{W Omega}, we can explicitly calculate the terms in the right hand side of \eqref{recall expansion U_2}. The first term becomes
\begin{align}
& \frac{d_N^2}{N} \left \langle \Big( W(\sqrt{N} \varphi) \U_2^{\dag}(t) a_y \U_2(t) \Omega \Big)^{(N-1)}, \Big( W(\sqrt{N} \varphi) \U_2^{\dag}(t) a_x \U_2(t) \Omega \Big)^{(N-1)} \right\rangle \nonumber \\
&= \frac{1}{N^2} \; \Big \langle \sum_{i=1}^{N-1} G_2(t, y, x_i) \prod_{j \neq i}^{N-1} \varphi(x_j) - N \int dz \; G_2 (t, y, z) \overline{\varphi}(z) \prod_{j=1}^{N-1} \varphi(x_j), \label{E_2 first term calculation} \\
& \qquad \qquad \sum_{i=1}^{N-1} G_2(t, x, x_i) \prod_{j \neq i}^{N-1} \varphi(x_j) - N \int dz \; G_2 (t, x, z) \overline{\varphi}(z) \prod_{j=1}^{N-1} \varphi(x_j) \Big \rangle_{L^2_{X_N}} \nonumber \\
&= \frac{N-1}{N^2} \int dz \; \overline{G_2}(t, y, z) G_2(t, x, z) - \frac{N-2}{N^2} \int dz \; \overline{G_2}(t, y, z) \varphi(z) \int dz\; G_2(t, x, z) \overline{\varphi}(z). \nonumber
\end{align}
Here $\langle \cdot \rangle_{L^2_{X_N}}$ denotes the inner product with respect to the variables $x_1, x_2, \cdots, x_N$ only. For the second term, we have
\begin{align}
& \overline{\varphi}_t (y) \frac{d_N^2}{\sqrt{N}} \left \langle \Big( W(\sqrt{N} \varphi) \Omega \Big)^{(N-1)},
\Big( W(\sqrt{N} \varphi) \U_2^{\dag}(t) a_x \U_2(t) \Omega \Big)^{(N-1)} \right \rangle \nonumber \\
&= \frac{\overline{\varphi}_t (y)}{N} \left \langle \prod_{j=1}^{N-1} \varphi(x_j), \sum_{i=1}^{N-1} G_2(t, x, x_i) \prod_{j \neq i}^{N-1} \varphi(x_j) - N \int dz \; G_2(t, x, z) \overline{\varphi}(z) \prod_{j=1}^{N-1}\varphi(x_j) \right \rangle_{L^2_{X_N}} \nonumber \\
&= \frac{\overline{\varphi}_t(y)}{N} \left( (N-1) \int dz \; G_2 (t, x, z) \overline{\varphi}(z) - N \int dz \; G_2 (t, x, z) \overline{\varphi}(z) \right) \label{E_2 second term calculation} \\
&= - \frac{\overline{\varphi}_t(y)}{N} \int dz \; G_2 (t, x, z) \overline{\varphi}(z). \nonumber
\end{align}
A similar calculation shows for the third term that
\begin{align}
& \varphi_t(x) \frac{d_N^2}{\sqrt{N}} \left \langle \Big( W(\sqrt{N} \varphi) \U_2^{\dag}(t) a_y \U_2(t) \Omega \Big)^{(N-1)}, \Big( W(\sqrt{N} \varphi) \Omega \Big)^{(N-1)} \right \rangle \nonumber \\
&= - \frac{\varphi_t(x)}{N} \int dz \; \overline{G_2}(t, y, z) \varphi(z). \label{E_2 third term calculation}
\end{align}

Therefore, the proposition follows from \eqref{E_2 first term calculation}, \eqref{E_2 second term calculation}, \eqref{E_2 third term calculation}, and Lemma \ref{G_2 bound}.

\section{Strichartz Type Estimates for $\U_2(t)$}

In this section, we prove Proposition \ref{L6 norm estimate} using Strichartz estimates. For the standard Strichartz estimate used in this section, see Theorem 2.3.3 of \cite{C}.

\subsection{Regularity from $\U_2(t)$ dynamics}

We will use the notation
\begin{equation}
H_0 := \int dx \nabla_x a_x^{\dag} \nabla_x a_x \label{kinetic energy}
\end{equation}
to denote the kinetic energy.

\begin{lem} \label{strichartz 1}
Let $H_0$ be defined in \eqref{kinetic energy}. Define an operator $\T$ on $\F_X$ by
\begin{equation}
\T(t) := e^{-it H_0}.
\end{equation}
Then, there exists a constant $C$ such that, for all $\psi \in \F_X$,
\begin{equation}
\| \T(\cdot) \psi \|_{\widetilde{\F}}^2 \leq C \| \psi \|^2.
\end{equation}
\end{lem}

\textit{Proof.}
It is trivial that 
\begin{equation}
\| e^{-it H_0} \psi^{(n)} (X_n) \|_{L_t^{\infty} L_{X_n}^2}^2 = \| \psi^{(n)} (X_n) \|_{L_{X_n}^2}^2. \label{trivial estimate 1}
\end{equation}
For any fixed $i$ and $j$, $1 \leq i < j \leq n$, let $\eta = x_i - x_j$. Applying the standard Strichartz estimate for Schr\"odinger operator on $\psi^{(n)}$, we obtain
\begin{align}
& \| e^{-it H_0} \psi^{(n)} (X_n) \|_{L_t^2 L_{x_i - x_j}^6 L_{x_i + x_j}^2 L_{X_n \backslash \{x_i, x_j \} }^2 }^2 = \| e^{-2it \Delta_{\eta}} \psi^{(n)} (X_n) \|_{L_t^2 L_{\eta}^6 L_{x_i + x_j}^2 L_{X_n \backslash \{x_i, x_j \} }^2}^2 \label{strichartz estimate 1} \\
& \leq C \| \psi^{(n)} (X_n) \|_{L_{\eta}^2 L_{x_i + x_j}^2 L_{X_n \backslash \{x_i, x_j \} }^2}^2 = C \| \psi^{(n)} (X_n) \|_{L_{X_n}^2}^2. \nonumber
\end{align}
Summing \eqref{trivial estimate 1} and \eqref{strichartz estimate 1} over $n$, we get the desired lemma.

\begin{lem} \label{strichartz 2}
Let $H_0$ be defined in \eqref{kinetic energy}. Define an operator $\Phi$ on $\F$ through
\begin{equation}
\Phi_{\psi}(t, X_n) = \int_0^t ds \: e^{-i(t-s) H_0} \psi(s, X_n)
\end{equation}
for any $\psi \in \F$. Then, there exists a constant $C$ such that, for all $\psi \in \F$,
\begin{equation}
\| \Phi_{\psi} \|_{\widetilde{\F}} \leq C T \| \psi \|_{\F}.
\end{equation}
\end{lem}

\textit{Proof.}
To prove the desired lemma, it suffices to show that
\begin{equation}
\|\Phi_{\psi}^{(n)} \|_{F_n} \leq C T \| \psi^{(n)} \|_{\F}. \label{strichartz in a sector}
\end{equation}
Recall that we only consider a fixed time interval $[0, T]$. From the standard Strichartz estimate, we have
\begin{equation}
\Big \| \int_0^t ds \: e^{-i(t-s) H_0} \psi^{(n)}(s, X_n) \Big \|_{L_t^{\infty} L_{X_n}^2} \leq C \| \psi^{(n)} (t, X_n) \|_{L_t^1 L_{X_n}^2}. \label{strichartz estimate 2}
\end{equation}

For any $g(t, X_n) \in L_t^2 L_{x_i-x_j}^{6/5} L_{x_i + x_j}^2 L_{X_n \backslash \{x_i, x_j \} }^2$, we obtain as the dual inequality of Lemma \ref{strichartz 1} that
\begin{equation}
\left \| \int_0^T dt \: e^{it H_0} g(t, X_n) \right \|_{L_{X_n}^2} \leq C \| g(t, X_n) \|_{L_t^2 L_{x_i-x_j}^{6/5} L_{x_i + x_j}^2 L_{X_n \backslash \{x_i, x_j \} }^2}. \label{dual strichartz 2}
\end{equation}
From the standard Strichartz estimate, we also have
\begin{equation}
\left \| \int_0^T ds \: e^{is H_0} \psi^{(n)}(s, X_n) \right \|_{L_{X_n}^2} \leq C \| \psi^{(n)} (t, X_n) \|_{L_t^1 L_{X_n}^2}. \label{dual strichartz 3}
\end{equation}
Combining \eqref{dual strichartz 2} and \eqref{dual strichartz 3}, we get
\begin{align}
& \left \| \int_0^T ds \: e^{-i(t-s) H_0} \psi^{(n)}(s, X_n) \right \|_{L_t^2 L_{x_i - x_j}^6 L_{x_i + x_j}^2 L_{X_n \backslash \{x_i, x_j \} }^2 } \nonumber \\
&= \sup_{g} \int_0^T dt \int_0^T ds \left \langle g(t, X_n), e^{-i(t-s) H_0} \psi^{(n)}(s, X_n) \right \rangle \label{strichartz estimate 3} \\
&= \sup_{g} \left \langle \int_0^T dt \: e^{it H_0} g(t, X_n), \int_0^T ds \: e^{is H_0} \psi^{(n)}(s, X_n) \right \rangle \nonumber \\
&\leq C \sup_{g} \| g(t, X_n) \|_{L_t^2 L_{x_i-x_j}^{6/5} L_{x_i + x_j}^2 L_{X_n \backslash \{x_i, x_j \} }^2} \| \psi^{(n)} (t, X_n) \|_{L_t^1 L_{X_n}^2} = C \| \psi^{(n)} (t, X_n) \|_{L_t^1 L_{X_n}^2} \nonumber,
\end{align}
where the supremum is taken over all funtions $g$ satisfying
\begin{equation}
\| g(t, X_n) \|_{L_t^2 L_{x_i-x_j}^{6/5} L_{x_i + x_j}^2 L_{X_n \backslash \{x_i, x_j \} }^2} = 1.
\end{equation}
Applying Christ-Kiselev lemma to \eqref{strichartz estimate 3}, we finally obtain that
\begin{equation}
\left \| \int_0^t ds \: e^{-i(t-s) H_0} \psi^{(n)}(s, X_n) \right \|_{L_t^2 L_{x_i - x_j}^6 L_{x_i + x_j}^2 L_{X_n \backslash \{x_i, x_j \} }^2 } \leq C \| \psi^{(n)} (t, X_n) \|_{L_t^1 L_{X_n}^2}. \label{strichartz estimate 4}
\end{equation}

Thus, it follows from \eqref{strichartz estimate 2} and \eqref{strichartz estimate 4} that
\begin{align}
\|\Phi_{\psi}^{(n)} \|_{F_n} \leq C \| \psi^{(n)} (t, X_n) \|_{L_t^1 L_{X_n}^2} \leq C T \| \psi^{(n)} (t, X_n) \|_{L_t^{\infty} L_{X_n}^2},
\end{align}
which proves the claim \eqref{strichartz in a sector}. This concludes the proof of the desired lemma.

\subsection{Estimate on $H_2(t)$}

To prove Proposition \ref{L6 norm estimate}, we need the following lemma.
\begin{lem} \label{H_2 estimate}
Let $V = V_1 + V_2$ with $V_1 \in L^3$ and $V_2 \in L^{\infty}$. Then, for all $j \in \mathbb{N}$, there exists a constant $C$ depending only on $\| V_1 \|_{L^3}$, $\| V_2 \|_{L^{\infty}}$ and $\| \varphi \|_{H^1}$, such that
\begin{equation}
\| \N^j (H_2(t) - H_0 ) \psi \| \leq C \| (\N+2)^{j+1} \psi \|
\end{equation}
for all $\psi \in \F_X$.
\end{lem}

\textit{Proof.}
Let 
\begin{align}
L &:= \int dx (V * |\varphi_t |^2 )(x) a_x^{\dag} a_x, \\
M &:= \iint dx dy V(x-y) \overline{\varphi_t} (x) \varphi_t (y) a_y^{\dag} a_x, \\
B &:= \frac{1}{2} \iint dx dy V(x-y) \overline{\varphi_t} (x) \overline{\varphi_t} (y) a_x a_y.
\end{align}
Then, $H_2 - H_0 = L + M + B + B^{\dag}$.

Note that
\begin{equation}
\big \| |V|*|\varphi|^2 \big \|_{\infty} \leq \|V_1\|_{L^3} \| \varphi \|_{L^6} \| \varphi \|_{L^2} + \|V_2\|_{L^{\infty}} \| \varphi \|_{L^2}^2 \leq C.
\end{equation}
Since $L$ commutes with $\N$ and $V * |\varphi_t |^2 \in L^{\infty}$, there exists a constant $C$ such that
\begin{equation}
\| \N^j L \psi \| = \| L \N^j  \psi \| \leq C \| \N^{j+1} \psi \|. \label{L estimate}
\end{equation}

To prove an estimate for $\| \N^j  M \psi \| = \| M \N^j \psi \|$, we observe that, for any $\xi \in \F_X$,
\begin{align}
& | \langle \xi, M \N^j \psi \rangle | = \left| \iint dx dy \; V(x-y) \overline{\varphi}_t (x) \varphi_t (y) \langle (\N + 1)^{-1/2} a_y \xi, (\N +1)^{1/2} a_x \N^j \psi \rangle \right| \nonumber \\
&\leq \big \| |V|*|\varphi|^2 \big \|_{\infty} \left( \int dy \| (\N + 1)^{-1/2} a_y \xi \|^2 \right)^{1/2} \left( \int dx \| (\N +1)^{1/2} a_x \N^j \psi \|^2 \right)^{1/2} \\
& \leq C \| \xi \| \| \N^{j+1} \psi \|. \nonumber
\end{align}
Since $\xi$ was arbitrary, this shows that
\begin{equation}
\| \N^j M \psi \| \leq C \| \N^{j+1} \psi \|. \label{M estimate}
\end{equation}

Estimates for $\| \N^j  B \psi \|$ and $\| \N^j  B^{\dag} \psi \|$ can be obtained similarly. Since
\begin{equation}
\iint dx dy \; |V(x-y)|^2 |\varphi_t (x)|^2 |\varphi_t (y)|^2 \leq \|V_1 \|_{L^3}^2 \| \varphi_t \|_{L^6}^2 \| \varphi_t \|_{L^2}^2 + \|V_2 \|_{L^{\infty}}^2 \| \varphi_t \|_{L^2}^2 \| \varphi_t \|_{L^2}^2 \leq C,
\end{equation}
we have that, for any $\xi \in \F_X$,
\begin{align}
& | \langle \xi, \N^j B \psi \rangle | = \left| \iint dx dy \; V(x-y) \overline{\varphi}_t (x) \varphi_t (y) \langle \xi, a_x a_y (\N-2)^j \psi \rangle \right| \nonumber \\
&\leq \left( \iint dx dy \; |V(x-y)|^2 |\varphi_t (x)|^2 |\varphi_t (y)|^2 \| \xi \|^2 \right)^{1/2} \left( \int dx dy \| a_x a_y (\N-2)^j \psi \|^2 \right)^{1/2} \nonumber \\
& \leq C \| \xi \| \| \N^{j+2} \psi \|.
\end{align}
Again, since $\xi$ was arbitrary, this shows that
\begin{equation}
\| \N^j B \psi \| \leq C \| \N^{j+1} \psi \|. \label{B estimate 1}
\end{equation}
Similarly, we also have that
\begin{equation}
\| \N^j B^{\dag} \psi \| \leq C \| (\N+2)^{j+1} \psi \|. \label{B estimate 2}
\end{equation}

Thus, from \eqref{L estimate}, \eqref{M estimate}, \eqref{B estimate 1}, and \eqref{B estimate 2}, we get
\begin{equation}
\| \N^j  (H_2 (t)- H_0) \psi \| \leq \| \N^j  L \psi \| + \| \N^j M\psi \| + \| \N^j (B + B^{\dag}) \psi \| \leq C\|(N+2)^{j+1}\psi\|,
\end{equation}
which proves the desired lemma.

\subsection{Proof of Proposition \ref{L6 norm estimate}}

From Lemma \ref{strichartz 1}, Lemma \ref{strichartz 2}, and Lemma \ref{H_2 estimate}, we can prove Proposition \ref{L6 norm estimate}.

\textit{Proof of Proposition \ref{L6 norm estimate}.}
By using the group $e^{-itH_0}$ generated by $H_0$, we can write the evolution of $\N^j \psi_2(t)$ as
\begin{equation}
\N^j \psi_2(t) = e^{-itH_0} \N^j \psi + \int_s^t d\sigma \: e^{-i(t-\sigma) H_0} \N^j (H_2(\sigma)-H_0) \psi_2(\sigma).
\end{equation}
Thus, by Lemma \ref{strichartz 1}, Lemma \ref{strichartz 2}, and Lemma \ref{H_2 estimate}, we have
\begin{align}
\| \N^j \psi_2 \|_{\widetilde{\F}} &\leq C \| \N^j \psi \| + C T \| \N^j (H_2(\cdot)-H_0) \psi_2(\cdot) \|_{\F} \\
&\leq C \| \N^j \psi \| + C T \| (\N+2)^{j+1} \psi_2 \|_{\F}, \nonumber
\end{align}
which was to be proved.

\section{Difference between $\U(t)$ Dynamics and $\U_2(t)$ Dynamics}

\subsection{Estimates on $H_3 (t)$ and $H_4$}

From the regularity we have seen in Proposition \ref{L6 norm estimate}, we can obtain the following estimates.

\begin{lem} \label{H_3 estimate}
Let $V = V_1 + V_2$ with $V_1 \in L^3$ and $V_2 \in L^{\infty}$. Then, for all $j \in \mathbb{N}$, there exists a constant $C$ depending only on $j$, $\| V_1 \|_{L^3}$, $\| V_2 \|_{L^{\infty}}$ and $\| \varphi \|_{H^1}$, such that
\begin{equation}
\| \N^j H_3(t) \psi \| \leq \frac{C}{\sqrt{N}} \| (\N+1)^{j+\frac{3}{2}} \psi \|
\end{equation}
for all $\psi \in \F_X$.
\end{lem}

\textit{Proof.}
Let
\begin{equation}
A_3 (t) = \iint dx dy V(x-y) \overline{\varphi}_t (y) a_x^{\dag} a_y a_x.
\end{equation}
Then,
\begin{equation}
\N^j H_3 (t) = \frac{1}{\sqrt N} \big( \N^j A_3^{\dag} (t) + \N^j A_3 (t) \big) \label{H_3 expansion}.
\end{equation}

Now we estimate $\N^j A_3^{\dag}(t)$ and $\N^j A_3 (t)$ separately. The first term $\N^j A_3^{\dag}(t)$ satisfies
\begin{align}
& \| \N^j A_3^{\dag}(t) \psi(X_n) \|_{L_{X_n}^2}^2 \nonumber \\
&= \sum_{n=3}^{\infty} n^{j+1} (n-1)^2 \int dx_3 \cdots dx_n \int dx_1 dx_2 |V(x_1 - x_2)|^2 |\varphi_t (x_2)|^2 |\psi^{(n-1)}(x_1, x_3, \cdots , x_n )|^2 \nonumber \\
& \quad + \: 2^{j+1} \int dx_1 dx_2 |V(x_1 - x_2)|^2 |\varphi_t (x_2)|^2 |\psi^{(1)}(x_1)|^2 \label{A_3 estimate 1} \\
&\leq 2 \sum_{n=3}^{\infty} n^{j+1} (n-1)^2 \nonumber \\
& \qquad \times \int dx_3 \cdots dx_n \big( \| V_1 \|_{L^3}^2 \|\varphi_t\|_{L^6}^2 + \| V_2 \|_{L^{\infty}}^2 \|\varphi_t\|_{L^2}^2 \big) \|\psi^{(n-1)} (x_1, x_3, \cdots, x_n)\|_{L_{x_1}^2}^2 \nonumber \\
& \quad + \: 2^{j+2} \big( \| V_1 \|_{L^3}^2 \|\varphi_t\|_{L^6}^2 + \| V_2 \|_{L^{\infty}}^2 \|\varphi_t\|_{L^2}^2 \big) \| \psi^{(1)} (x_1)\|_{L_{x_1}^2}^2 \nonumber \\
&\leq C \| (\N+1)^{j+\frac{3}{2}} \psi \|^2, \nonumber
\end{align}
and $\N^j A_3 (t)$ satisfies
\begin{align}
& \| \N^j A_3(t) \psi(X_n) \|_{L_{X_n}^2}^2 \nonumber \\
&= \sum_{n=1}^{\infty} n^{j+2} (n+1) \int dx_1 \cdots dx_n \Big| \int dy \: V(x_1 - y) \varphi_t (y) \psi^{(n+1)}(y, x_1, \cdots , x_n ) \Big|^2 \label{A_3 estimate 2} \\
&\leq 2 \sum_{n=1}^{\infty} n^{j+2} (n+1) \int dx_1 \cdots dx_n \big( \| V_1 \|_{L^3}^2 \|\varphi_t\|_{L^6}^2 + \| V_2 \|_{L^{\infty}}^2 \|\varphi_t\|_{L^2}^2 \big) \| \psi^{(n+1)} (y, x_1, \cdots, x_n)\|_{L_y^2}^2 \nonumber \\
&\leq C \| \N^{j+\frac{3}{2}} \psi \|^2. \nonumber
\end{align}

Hence, from \eqref{H_3 expansion}, \eqref{A_3 estimate 1}, and \eqref{A_3 estimate 2} we get
\begin{equation}
\| \N^j H_3(t) \psi \| \leq \frac{C}{\sqrt N} \| (\N+1)^{j+\frac{3}{2}} \psi \|,
\end{equation}
which was to be proved.

\begin{lem} \label{H_4 estimate}
Let $V = V_1 + V_2$ with $V_1 \in L^3$ and $V_2 \in L^{\infty}$. Then, for all $j \in \mathbb{N}$, there exists a constant $C$ depending only on $j$, $\| V_1 \|_{L^3}$, $\| V_2 \|_{L^{\infty}}$ and $\| \varphi \|_{H^1}$, such that
\begin{equation}
\int_0^t ds \| \N^j H_4 \U_2 (s;s_1) \psi \|^2 \leq \frac{C}{N^2} \| \N^{j+2} \psi \|^2 + \frac{C T}{N^2} \| (\N+2)^{j+3} \U_2 (\cdot ;s_1) \psi \|_{\F}^2
\end{equation}
for all $\psi \in \F_X$.
\end{lem}

\textit{Proof.}
From H\"older's inequality, we have
\begin{align}
& \| \N^j H_4 \U_2 (s;s_1) \psi \|^2 = \frac{1}{N^2} \sum_{n=2}^{\infty} n^{2j} \int dX_n \Big| \sum_{i<j}^n V(x_i-x_j) \big( \U_2 (s;s_1) \psi \big)^{(n)} (X_n) \Big|^2 \nonumber \\
& \leq \frac{1}{N^2} \sum_{n=2}^{\infty} n^{2j} \frac{n(n-1)}{2} \sum_{i<j}^n \int dX_n \Big| V(x_i-x_j) \big( \U_2 (s;s_1) \psi \big)^{(n)} (X_n) \Big|^2 \nonumber \\
& \leq \frac{C}{N^2} \sum_{n=2}^{\infty} n^{2j+2} \sum_{i<j}^n \int d(x_i - x_j) |V(x_i-x_j)|^2 \\
& \qquad \qquad \times \int d(x_i + x_j) dx_1 \cdots \widehat{dx_i} \cdots \widehat{dx_j} \cdots dx_n \Big| \big( \U_2 (s;s_1) \psi \big)^{(n)} (x_1, \cdots, x_n) \Big|^2 \nonumber \\
& \leq \frac{C}{N^2} \sum_{n=2}^{\infty} n^{2j+2} \sum_{i<j}^n \Big( \| V_1 \|_{L^3}^2 \| \big( \U_2 (s;s_1) \psi \big)^{(n)} (x_1, \cdots, x_n) \|_{L_{x_i-x_j}^6 L_{x_i+x_j}^2 L_{X_n \backslash \{x_i, x_j\}}^2}^2 \nonumber \\
& \qquad \qquad \qquad \qquad + \| V_2 \|_{L^{\infty}}^2 \| \big( \U_2 (s;s_1) \psi \big)^{(n)} (x_1, \cdots, x_n) \|_{L_{X_n}^2}^2 \Big). \nonumber
\end{align}
Thus,
\begin{equation}
\int_0^t ds \| \N^j H_4 \U_2 (s;s_1) \psi \|^2 \leq \frac{C}{N^2} \| \N^{j+2} \U_2(\cdot;s_1) \psi \|_{\widetilde{\F}}^2 + \frac{CT}{N^2} \| \N^{j+2} \U_2(\cdot;s_1) \psi \|_{\F}^2.
\end{equation}
Proposition \ref{L6 norm estimate} shows that
\begin{equation}
\| \N^{j+2} \U_2(\cdot;s_1) \psi \|_{\widetilde{\F}} \leq C \|\N^{j+2} \psi\| + C T \| (\N+2)^{j+2} \U_2(\cdot;s_1) \psi \|_{\F}.
\end{equation}
Therefore,
\begin{equation}
\int_0^t ds \| \N^j H_4 \U_2 (s;s_1) \psi \|^2 \leq \frac{C}{N^2} \| \N^{j+2} \psi \|^2 + \frac{C T}{N^2} \| (\N+2)^{j+3} \U_2 (\cdot;s_1) \psi \|_{\F}^2,
\end{equation}
which was to be proved.

\subsection{Proof of Proposition \ref{R_y estimate}}

We are ready to estimate the difference between $\U^{\dag} (t) a_y \U(t) \Omega$ and $\U_2^{\dag} (t) a_y \U_2(t) \Omega$ by proving Proposition \ref{R_y estimate}. Note that we are free to use the Duhamel formula with the operators $\U$ and $\U_2$ on a dense subset $\D (H_0) \cap \D (\N)$ of $\F_X$, which can be seen from equations \eqref{evolution of U} and \eqref{evolution of U_2} together with the fact that the Weyl operator $W( \sqrt N \varphi_s )$ maps $\D (H_0)$ onto $\D(H_0)$ if $\varphi_s \in H^1 (\mathbb{R}^3)$. (See also Lemma 3.1 of \cite{GV1}.)

\textit{Proof.}
Let
\begin{equation}
\R_y^1 (t) := \big( \U^{\dag} (t) - \U_2^{\dag}(t) \big) a_y \U(t)
\end{equation}
and
\begin{equation}
\R_y^2 (t) := \U_2^{\dag} (t) a_y \big( \U(t) - \U_2(t) \big).
\end{equation}
Then, $\R_y(t) = \R_y^1 (t) + \R_y^2 (t)$.

Since
\begin{equation}
\U^{\dag}(t) - \U_2^{\dag}(t) = i \int_0^t ds \: \U^{\dag}(s;0) (H_3(s) + H_4) \U_2^{\dag}(t;s),
\end{equation}
applying $\R_y^1 (t)$ on the vacuum gives
\begin{align}
& \| (\N+1)^{\frac{j}{2}} \R_y^1 (t) \Omega \| = \| (\N+1)^{\frac{j}{2}} (\U^{\dag}(t)-\U_2^{\dag}(t)) a_y \U(t) \Omega \| \nonumber \\
&= \Big \| \int_0^t ds (\N+1)^{\frac{j}{2}} \U^{\dag}(s) (H_3(s)+H_4) \U_2(s;t) a_y \U(t) \Omega \Big \| \\
&\leq \int_0^t ds \| (\N+1)^{\frac{j}{2}} \U^{\dag}(s) (H_3(s)+H_4) \U_2(s;t) a_y \U(t) \Omega \| \nonumber \\
&= \int_0^t ds \langle \U^{\dag}(s) (H_3(s)+H_4) \U_2(s;t) a_y \U(t) \Omega, (\N+1)^j \U^{\dag}(s) (H_3(s)+H_4)
\U_2(s;t) a_y \U(t) \Omega \rangle^{\frac{1}{2}}. \nonumber
\end{align}
Thus, from Lemma \ref{U conserves N} with Schwarz inequality, we obtain that
\begin{align}
& \int dy \| (\N+1)^{\frac{j}{2}} \R_y^1 (t) \Omega \|^2 \nonumber \\
& \leq Ct \int dy \int_0^t ds \|(\N+1)^{j+1} H_3(s) \U_2(s;t) a_y \U(t) \Omega \|^2 \label{R1 estimate} \\
& \quad + Ct \int dy \int_0^t ds \|(\N+1)^{j+1} H_4 \U_2(s;t) a_y \U(t)\Omega\|^2. \nonumber
\end{align}

Now, from Lemma \ref{U conserves N}, Lemma \ref{U_2 conserves N}, and Lemma \ref{H_3 estimate}, the first term in the right hand side of \eqref{R1 estimate} can be estimated as
\begin{align}
& \int dy \int_0^t ds \|(\N+1)^{j+1} H_3(s) \U_2(s;t) a_y \U(t) \Omega \|^2 \nonumber \\
& \leq \frac{C}{N} \int dy \int_0^t ds \| (\N+1)^{j+\frac{5}{2}} \U_2(s;t) a_y \U(t) \Omega \|^2 \nonumber \\
& \leq \frac{C e^{KT}}{N} \int dy \int_0^t ds \langle a_y \U(t) \Omega, (\N+1)^{2j+5} a_y \U(t) \Omega \rangle \label{R1-H_3 estimate} \\
& \leq \frac{C e^{KT}}{N} \int_0^t ds \langle \U(t) \Omega, (\N+1)^{2j+6} \U(t) \Omega \rangle \leq \frac{C e^{KT}}{N} \int_0^t ds \langle \Omega, (\N+1)^{4j+14} \Omega \rangle \nonumber \\
& \leq \frac{C e^{KT}}{N}. \nonumber
\end{align}
Similarly, from Lemma \ref{U conserves N}, Lemma \ref{U_2 conserves N}, and Lemma \ref{H_4 estimate}, the second term in the right hand side of \eqref{R1 estimate} can be estimated as
\begin{align}
& \int dy \int_0^t ds \|(\N+1)^{j+1} H_4 \U_2(s;t) a_y \U(t) \Omega \|^2 \nonumber \\
& \leq \frac{C}{N^2} \int dy \| (\N+1)^{j+3} a_y \U(t) \Omega \|^2 + \frac{C T}{N^2} \int dy \| (\N+1)^{j+4} \U_2(\cdot;t) a_y \U(t) \Omega \|_{\F}^2 \nonumber \\
& \leq \frac{C}{N^2} \langle \U(t) \Omega, \N^{2j+7} \U(t) \Omega \rangle + \frac{C e^{KT}}{N^2} \langle \U(t) \Omega, \N^{2j+9} \U(t) \Omega \rangle \label{R1-H_4 estimate} \\
& \leq \frac{C e^{KT}}{N^2} \langle \Omega, (\N+1)^{4j+20} \Omega \rangle = \frac{C e^{KT}}{N^2}. \nonumber
\end{align}

Hence, from \eqref{R1 estimate}, \eqref{R1-H_3 estimate}, and \eqref{R1-H_4 estimate}, we get
\begin{equation}
\int dy \| (\N+1)^{\frac{j}{2}} \R_y^1(t) \Omega \|^2 \leq \frac{C e^{KT}}{N}.
\end{equation}
The study of $\R_y^2 (t)$ is similar and gives
\begin{equation}
\int dy \| (\N+1)^{\frac{j}{2}} \R_y^2(t) \Omega \|^2 \leq \frac{C e^{KT}}{N}.
\end{equation}
Therefore,
\begin{equation}
\left( \int dy \| (\N+1)^{\frac{j}{2}} \R_y (t) \Omega \|^2 \right)^{\frac{1}{2}} \leq \frac{C e^{KT}}{\sqrt N},
\end{equation}
which was to be proved.

\section{Proof of Main Theorem}

Now, we are ready to prove the main theorem.

\textit{Proof of Main Theorem.}
First, we write the difference between $E(t, x, y)$ and $E_2(t, x, y)$ in detail as follows.
\begin{align}
& E(t, x, y)-E_2(t, x, y) \nonumber \\
&= \frac{d_N^2}{N} \left \langle \Big( W(\sqrt{N} \varphi) \R_y(t) \Omega \Big)^{(N-1)}, \Big( W(\sqrt{N} \varphi) \U_2^{\dag}(t) a_x \U_2(t) \Omega \Big)^{(N-1)} \right \rangle \nonumber \\
& \quad + \frac{d_N^2}{N} \left \langle \Big( W(\sqrt{N} \varphi) \U_2^{\dag}(t) a_y \U_2(t) \Omega \Big)^{(N-1)},
\Big(W(\sqrt{N} \varphi) \R_x(t) \Omega \Big)^{(N-1)} \right \rangle \label{difference between E and E_2} \\
& \quad + \frac{d_N^2}{N} \left \langle \Big( W(\sqrt{N} \varphi) \R_y(t) \Omega \Big)^{(N-1)}, \Big(W(\sqrt{N} \varphi) \R_x(t) \Omega \Big)^{(N-1)} \right \rangle \nonumber \\
& \quad + \overline{\varphi_t}(y) \frac{d_N^2}{\sqrt{N}} \left \langle \Big( W(\sqrt{N} \varphi) \Omega \Big)^{(N-1)},
\Big(W(\sqrt{N} \varphi) \R_x(t) \Omega \Big)^{(N-1)} \right \rangle \nonumber \\
& \quad + \varphi_t(x) \frac{d_N^2}{\sqrt{N}} \left \langle \Big( W(\sqrt{N} \varphi) \R_y(t) \Omega \Big)^{(N-1)}, \Big(W(\sqrt{N} \varphi) \Omega \Big)^{(N-1)} \right \rangle. \nonumber
\end{align}

From Lemma \ref{U_2 conserves N} and Proposition \ref{R_y estimate}, the first term in the right hand side of \eqref{difference between E and E_2} can be estimate as
\begin{align}
& \frac{d_N^2}{N} \left( \int dx dy \left| \left \langle \Big( W(\sqrt{N} \varphi) \R_y(t) \Omega \Big)^{(N-1)}, \Big( W(\sqrt{N} \varphi) \U_2^{\dag}(t) a_x \U_2(t) \Omega \Big)^{(N-1)} \right \rangle \right|^2 \right)^{\frac{1}{2}} \nonumber \\
& \leq \frac{d_N^2}{N} \left( \int dy \| \R_y(t) \Omega \|^2 \right)^{\frac{1}{2}} \left( \int dx \| \U_2^{\dag}(t) a_x \U_2(t) \Omega \|^2 \right)^{\frac{1}{2}} \label{first term} \\
& \leq \frac{d_N^2}{N} \cdot \frac{C e^{KT}}{\sqrt N} \cdot C e^{KT} \leq \frac{C e^{KT}}{N}. \nonumber
\end{align}
Similarly, the second term can be estimated as
\begin{equation}
\frac{d_N^2}{N} \left( \int dx dy \left| \left \langle \Big( W(\sqrt{N} \varphi) \U_2^{\dag}(t) a_y \U_2(t) \Omega \Big)^{(N-1)}, \Big( W(\sqrt{N} \varphi) \R_x(t) \Omega \Big)^{(N-1)} \right \rangle \right|^2 \right)^{\frac{1}{2}} \leq \frac{C e^{KT}}{N}. \label{second term}
\end{equation}
To estimate the third term, we use the following estimate
\begin{align}
& \frac{d_N^2}{N} \left( \int dx dy \left| \left \langle \Big( W(\sqrt{N} \varphi) \R_y(t) \Omega \Big)^{(N-1)}, \Big(W(\sqrt{N} \varphi) \R_x(t) \Omega \Big)^{(N-1)} \right \rangle \right|^2 \right)^{\frac{1}{2}} \nonumber \\
& \leq \frac{d_N^2}{N} \left( \int dy \| \R_y(t) \Omega \|^2 \right)^{\frac{1}{2}} \left( \int dx \| \R_x(t) \Omega \|^2 \right)^{\frac{1}{2}} \leq \frac{C e^{KT}}{N \sqrt N}. \label{third term}
\end{align}

To estimate the fourth term in the right hand side of \eqref{difference between E and E_2}, we first note that
\begin{align}
& \left \langle \Big( W(\sqrt{N} \varphi) \Omega \Big)^{(N-1)}, \Big(W(\sqrt{N} \varphi) \R_x(t) \Omega \Big)^{(N-1)} \right \rangle \nonumber \\
&= \frac{1}{d_N} \left \langle \varphi^{\otimes (N-1)}, \Big(W(\sqrt{N} \varphi) \R_x(t) \Omega \Big)^{(N-1)} \right \rangle \nonumber \\
& = \frac{1}{d_N} \left \langle \varphi^{\otimes (N-1)}, W(\sqrt{N} \varphi) \R_x(t) \Omega \right \rangle = \frac{1}{d_N} \left \langle W(\sqrt{N} \varphi)^{\dag} \varphi^{\otimes (N-1)}, \R_x(t) \Omega \right \rangle \label{fourth term 1} \\
& = \frac{1}{d_N} \left \langle (\N+1)^{-\frac{1}{2}} W(\sqrt{N} \varphi)^{\dag} \varphi^{\otimes (N-1)}, (\N+1)^{\frac{1}{2}} \R_x(t) \Omega \right \rangle. \nonumber
\end{align}
It follows from Proposition \ref{R_y estimate} and Lemma \ref{W estimate for factorized data} that
\begin{align}
& \left( \int dx \left| \left \langle (\N+1)^{-\frac{1}{2}} W(\sqrt{N} \varphi)^{\dag} \varphi^{\otimes (N-1)}, (\N+1)^{\frac{1}{2}} \R_x(t) \Omega \right \rangle \right|^2 \right)^{\frac{1}{2}} \nonumber \\
& \leq \left \| (\N+1)^{-\frac{1}{2}} W(\sqrt{N} \varphi)^{\dag} \varphi^{\otimes (N-1)} \right \| \left( \int dx \left \| (\N+1)^{\frac{1}{2}} \R_x(t) \Omega \right \|^2 \right)^{\frac{1}{2}} \label{fourth term 2} \\
& \leq \frac{C e^{KT}}{d_N \sqrt{N}} \nonumber.
\end{align}
Thus, from \eqref{fourth term 1} and \eqref{fourth term 2}, we obtain
\begin{align}
& \left( \int dx dy \left| \overline{\varphi_t}(y) \frac{d_N^2}{\sqrt{N}} \left \langle \Big( W(\sqrt{N} \varphi) \Omega \Big)^{(N-1)}, \Big(W(\sqrt{N} \varphi) \R_x(t) \Omega \Big)^{(N-1)} \right \rangle \right|^2 \right)^{\frac{1}{2}} \label{fourth term} \\
& \leq \frac{C e^{KT}}{N}. \nonumber
\end{align}

The last term in the right hand side of \eqref{difference between E and E_2} can also be similarly estimated as
\begin{align}
& \left( \int dx dy \left| \varphi_t(x) \frac{d_N^2}{\sqrt{N}} \left \langle \Big( W(\sqrt{N} \varphi) \R_y(t) \Omega \Big)^{(N-1)}, \Big(W(\sqrt{N} \varphi) \Omega \Big)^{(N-1)} \right \rangle \right|^2 \right)^{\frac{1}{2}} \label{fifth term} \\
& \leq \frac{C e^{KT}}{N}. \nonumber
\end{align}

Therefore, together with Proposition \ref{E_2 bound}, inserting \eqref{first term}, \eqref{second term}, \eqref{third term}, \eqref{fourth term}, and, \eqref{fifth term} into \eqref{difference between E and E_2} yields
\begin{equation}
\| E(t, x, y) \|_{L_{x, y}^2} \leq \| E_2(t, x, y) \|_{L_{x, y}^2} + \frac{C e^{KT}}{N} \leq \frac{C e^{KT}}{N}.
\end{equation}
By Remark \ref{time interval} and Remark \ref{HS norm}, this completes the proof of the main theorem.

\section{Expectation of Number Operator with respect to the Various Evolutions}

The following lemma shows that the expectation of the number operator with respect to the full evolution is bounded uniformly in $N$.
\begin{lem}[Proposition 3.3 of \cite{RS}] \label{U conserves N}
Suppose that $V = V_1 + V_2$ where $V_1 \in L^3 (\mathbb{R}^3)$ and $V_2 \in L^{\infty} (\mathbb{R}^3 )$. Let $\U(t;s)$ be the operator satisfying \eqref{evolution of U}. Then, for all $j \in \mathbb{N}$, there exist constants $C$ and $K$, depending only on $j$, $\|\varphi\|_{H^1}$, and $\| V \|_{L_3}$ such that
\begin{equation}
\langle \U(t;s) \psi, \N^j \U(t;s) \psi \rangle \leq C e^{K|t-s|} \langle \psi, (\N+1)^{2j+2} \psi \rangle,
\end{equation}
for all $\psi \in \F_X$.
\end{lem}

\begin{remark}
Specially, if $\psi = \Omega$, we have
\begin{equation}
\langle \U(t;s) \Omega, \N^j \U(t;s) \Omega \rangle \leq C e^{K|t-s|}.
\end{equation}
\end{remark}

A similar result holds for $\U_2 (t)$ evolution. For completeness, we prove it here.
\begin{lem} \label{U_2 conserves N}
Suppose that $V = V_1 + V_2$ where $V_1 \in L^3 (\mathbb{R}^3)$ and $V_2 \in L^{\infty} (\mathbb{R}^3 )$. Let $\U_2(t;s)$ be the operator satisfying \eqref{evolution of U_2}. Then, for all $j \in \mathbb{N}$, there exist constants $C$ and $K$, depending only on $j$, $\|\varphi\|_{H^1}$, and $\| V \|_{L_3}$ such that
\begin{equation}
\langle \U_2(t;s) \psi, \N^j \U_2(t;s) \psi \rangle \leq C e^{K|t-s|} \langle \psi, (\N+1)^j \psi \rangle,
\end{equation}
for all $\psi \in \F_X$.
\end{lem}

\textit{Proof.}
It suffices to prove the lemma when $s=0$. We have
\begin{align}
& \frac{d}{dt}\langle \U_2(t;0)\psi, (\N+1)^j\U_2(t;0) \psi \rangle = \langle \U_2(t;0)\psi, [i H_2,(\N+1)^j] \U_2(t;0) \psi \rangle \nonumber \\
&= \text{Im} \iint dxdy V(x-y)\varphi_t(x)\varphi_t(y)\langle \U_2(t;0)\psi, [a_x^{\dag} a_y^{\dag}, (\N+1)^j] \U_2(t;0) \psi \rangle \nonumber \\
&= \text{Im} \sum^{j-1}_{k=0} \binom{j}{k} (-1)^k \iint dx dy V(x-y) \varphi_t(x) \varphi_t(y) \label{U_2 gronwall lemma} \\
& \; \times \Big(\langle a_x(\N+1)^{\frac{k}{2}} \U_2(t;0) \psi, a_y^{\dag}(\N+3)^\frac{k}{2} \U_2(t;0) \psi \rangle + \langle a_x \N^{\frac{k}{2}} \U_2(t;0) \psi, a_y^{\dag}(\N+2)^\frac{k}{2} \U_2(t;0) \psi \rangle \Big) \nonumber \\
&\leq \sum^{j-1}_{k=0} \binom{j}{k} \sup_x \| V(x-\cdot)\varphi(\cdot)\|_{L^2} \Big(\int dx |\varphi_t(x)| \|a_x(\N+1)^{\frac{k}{2}} \U_2(t;0) \psi \| \|(\N+3)^{\frac{k+1}{2}} \U_2(t;0) \psi \| \nonumber \\
& \qquad +\int dx |\varphi_t(x)| \| a_x \N^{\frac{k}{2}} \U_2(t;0) \psi \| \| (\N+3)^{\frac{k+1}{2}} \U_2(t;0) \psi \| \Big) \nonumber \\
&\leq C \|(\N+1)^{\frac{j}{2}} \U_2(t;0) \psi \|^2 = C\langle \U_2(t;0)\psi, (\N+1)^j \U_2(t;0) \psi \rangle. \nonumber
\end{align}
Since $\U_2(0;0)=I$, we also have
\begin{equation}
\langle \U_2(0;0) \psi, (\N+1)^j\U_2(0;0) \psi \rangle =\langle \psi, (\N+1)^j \psi \rangle. \label{U_2 gronwall initial}
\end{equation}
Using \eqref{U_2 gronwall lemma} and \eqref{U_2 gronwall initial} the conclusion follows directly from the Gronwall's lemma.

\begin{lem} \label{W estimate for factorized data}
There exists a constant $C$ such that
\begin{equation}
\Big\| (\N+1)^{-\frac{1}{2}} W(\sqrt{N} \varphi)^{\dag} (\varphi^{\otimes (N-1)}) \Big\| \leq \frac{C}{d_N}.
\end{equation}
\end{lem}

\textit{Proof.}
It is obvious that $W(\sqrt{N} \varphi)^{\dag} (\varphi^{\otimes (N-1)})$ is a linear combination of tensor products of $\varphi$. We define the coefficients $\A_m$ such that
\begin{align}
\A_m := \left\| \Big( W(\sqrt{N} \varphi)^{\dag} (\varphi^{\otimes (N-1)}) \Big)^{(m)} \right\|.
\end{align}
For $m \leq N-1$, we can explicitly calculate $\A_m$. Since
\begin{align}
\Big( W(\sqrt{N} \varphi)^{\dag} (\varphi^{\otimes (N-1)}) \Big)^{(m)} = \Big( e^{-N /2} \textrm{exp}(a^{\dag}(- \sqrt{N} \varphi)) \textrm{exp}( a(\sqrt{N} \varphi)) (\varphi^{\otimes (N-1)}) \Big)^{(m)},
\end{align}
we have
\begin{align}
\A_m = e^{-N/2} \sum_{k=0}^m &\frac{1}{k!} (- \sqrt{N})^k \sqrt{m-k+1} \sqrt{m-k+2} \cdots \sqrt{m} \\
& \quad \times \frac{1}{(N-m-1+k)!} \sqrt{N}^{N-m-1+k} \sqrt{N-1} \sqrt{N-2} \cdots \sqrt{m-k+1}. \nonumber
\end{align}

We want to estimate $|\A_m|$. When $m=0$, we have
\begin{equation}
\A_0 = e^{-N/2} \frac{\sqrt{N}^{N-1}}{\sqrt{(N-1)!}} = \frac{1}{d_N}. \label{A_m bound 1}
\end{equation}
When $1 \leq m \leq N-1$, $\A_m$ can be simplified as follows.
\begin{align}
& \A_m = e^{-N/2} \sqrt{N}^{N-m-1} \sqrt{\frac{(N-1)!}{m!}} \sum_{k=0}^m (-1)^k N^k \frac{m!}{(m-k)! k! (N-m-1+k)!} \\
&= e^{-N/2} \sqrt{N}^{N-m-1} \sqrt{\frac{m!}{(N-1)!}} L_m^{(N-m-1)}(N), \nonumber
\end{align}
where the associated Laguerre polynomial $L_n^{(\alpha)}(x)$ is defined by
\begin{equation}
L_n^{(\alpha)}(x) := \sum_{k=0}^n (-1)^k \frac{(n+\alpha)!}{k! (n-k)! (\alpha+k)!} x^k.
\end{equation}

We now follow the argument used in the proof of Lemma 4.2 in \cite{RS}. The sharp estimate obtained by Krasikov \cite{K} for $L_n^{(\alpha)}(x)$ shows that
\begin{equation}
| L_n^{(\alpha)}(x) | < \sqrt{\frac{(n+\alpha)!}{n!}} \sqrt{\frac{x(s^2-q^2)}{r(x)}} e^{\frac{x}{2}} x^{-\frac{\alpha+1}{2}},
\end{equation}
where
\begin{equation}
s = (n+\alpha+1)^{\frac{1}{2}} + n^{\frac{1}{2}}, \quad q = (n+\alpha+1)^{\frac{1}{2}} - n^{\frac{1}{2}}, \quad r(x) = (x-q^2)(s^2-x).
\end{equation}
Thus, we obtain that, for $1 \leq m \leq N-1$,
\begin{align}
|\A_m| &< e^{-N/2} \sqrt{N}^{N-m-1} \sqrt{\frac{m!}{(N-1)!}} \sqrt{\frac{(N-1)!}{m!}} \sqrt{\frac{4N \sqrt{Nm}}{4Nm - m^2}} e^{N/2} N^{-\frac{N-m}{2}} = \sqrt{\frac{4 \sqrt{Nm}}{4Nm - m^2}} \nonumber \\
&< C N^{-1/4} m^{-1/4}. \label{A_m bound 2}
\end{align}

For $m \geq N$, we only need the following bound.
\begin{equation}
\sum_{m=N}^{\infty} |\A_m|^2 \leq \left\| \Big( W(\sqrt{N} \varphi)^{\dag} (\varphi^{\otimes (N-1)}) \Big) \right\|^2 \leq 1. \label{A_m bound 3}
\end{equation}

Therefore, from \eqref{A_m bound 1}, \eqref{A_m bound 2}, and \eqref{A_m bound 3}, we obtain
\begin{align}
& \Big\| (\N+1)^{-\frac{1}{2}} W(\sqrt{N} \varphi)^{\dag} (\varphi^{\otimes (N-1)}) \Big\|^2 = \sum_{m=0}^{\infty} \frac{|\A_m|^2}{m+1}  \\
& \leq \frac{1}{d_N^2} + C N^{-1/2} \sum_{m=1}^{N-1} \frac{1}{(m+1)^{3/2}} + \frac{C}{N} \sum_{m=N}^{\infty} |\A_m|^2 \leq C N^{-1/2}. \nonumber
\end{align}
This proves the desired lemma.

\section{Stability of the Operator}

The following lemma shows that the operator $H_2 (t)$ is stable. (See Proposition 3.4 of \cite{Ka} for more details.)

\begin{lem} \label{stability}
There exists a constant $C, K>0$ such that, for the operator $A_2 (t) = H_2 (t) + C (\N +1)$, we have the operator inequality $\dot{A_2} (t) \leq K A_2 (t)$, where $\dot{A_2} (t) = (d/dt) A_2 (t)$.
\end{lem}

\textit{Proof.}
Let
\begin{align}
&\dot{H_2}(t) := \frac{d H_2 (t)}{dt} \nonumber \\
&= \iint dx dy V(x-y) \overline{\varphi_t (y)} \dot{\varphi_t}(y) a_x^{\dag} a_x + \iint dx dy V(x-y) \overline{\varphi_t (x)} \dot{\varphi_t}(y) a_y^{\dag} a_x \nonumber \\
& \quad + \iint dx dy V(x-y) \varphi_t (x) \dot{\varphi_t}(y) a_x^{\dag} a_y^{\dag} + h.c. \label{H2 derivative}
\end{align}
where h.c. denotes the Hermitian conjugate and $\dot{\varphi_t} = \partial_t \varphi_t$.

In order to control $\langle \psi, \dot{H_2}(t) \psi \rangle$ for $\psi \in \F_X$, we need to estimate terms such as
\begin{equation}
\iint dx dy \langle \psi, V(x-y) \varphi_t (x) \dot{\varphi_t}(y) a_x^{\dag} a_y^{\dag} \psi \rangle.
\end{equation}
We know that $\varphi_t$ is the solution of the nonlinear Hartree equation and satisfies
\begin{equation}
\dot{\varphi_t}(y) = -i [ -\Delta \varphi_t (y) + (V * |\varphi_t|^2 )(y) \varphi_t (y) ].
\end{equation}
Thus,
\begin{align}
& \iint dx dy \left \langle \psi, V(x-y) \varphi_t (x) \dot{\varphi_t}(y) a_x^{\dag} a_y^{\dag} \psi \right \rangle \nonumber \\
&= -i \iint dx dy \left \langle a_x a_y \psi, V(x-y) \varphi_t (x) [(-\Delta \varphi_t )(y) + (V * |\varphi_t|^2 )(y) \varphi_t (y) ] \psi \right \rangle. \label{H2 derivative expansion}
\end{align}

Since $\varphi_t \in H^1$ and
\begin{equation}
\| V * |\varphi_t|^2 \|_{\infty} \leq C,
\end{equation}
we obtain that
\begin{align}
& \Big| \iint dx dy \left \langle a_x a_y \psi, V(x-y) \varphi_t (x) (V * |\varphi_t|^2 )(y) \varphi_t (y) \psi \right \rangle \Big| \nonumber \\
&\leq \sum_{n=2}^{\infty} \sqrt{n(n-1)} \| \psi^{(n-2)} \| \cdot \| \psi^{(n)} \| \cdot \| V(x-y) \varphi_t (x) (V * |\varphi_t|^2 )(y) \varphi_t (y) \|_{L_{x, y}^2} \label{H2 derivative nonlinear estimate} \\
&\leq C \langle \psi, \N \psi \rangle. \nonumber
\end{align}

By integrating by parts, we find that
\begin{align}
& \iint dx dy \; \overline{\psi^{(n)} (x, y, X_{n-2})} V(x-y) \varphi_t (x) (-\Delta \varphi_t) (y) \nonumber \\
&= \iint dx dy \left( \overline{(\nabla_y \psi^{(n)}) (x, y, X_{n-2})} V(x-y) - \overline{\psi^{(n)} (x, y, X_{n-2})} (\nabla V)(x-y) \right) \cdot \varphi_t (x) \nabla \varphi_t (y) \nonumber \\
&= \iint dx dy \; \overline{(\nabla_y \psi^{(n)}) (x, y, X_{n-2})} \cdot V(x-y) \varphi_t (x) \nabla \varphi_t (y) \\
& \quad + \iint dx dy \; \overline{(-\nabla_x \psi^{(n)}) (x, y, X_{n-2})} \cdot V(x-y) \varphi_t (x) \nabla \varphi_t (y) \nonumber \\
& \quad + \iint dx dy \; \overline{\psi^{(n)} (x, y, X_{n-2})} V(x-y) (-\nabla \varphi_t) (x) \cdot \nabla \varphi_t (y). \nonumber
\end{align}
Thus, we get
\begin{align}
& \Big| \iint dx dy \left \langle a_x a_y \psi, V(x-y) \varphi_t (x) (-\Delta \varphi_t) (y) \psi \right \rangle \Big| \label{H2 derivative linear term 1} \\
&= \Big| \sum_{n=2}^{\infty} \sqrt{n(n-1)} \int dx dy dX_{n-2} \; \overline{\psi^{(n)} (x, y, X_{n-2})} V(x-y) \varphi_t (x) (-\Delta \varphi_t) (y) \psi^{(n-2)} (X_{n-2}) \Big| \nonumber \\
&\leq \sum_{n=2}^{\infty} \sqrt{n(n-1)} \nonumber \\
& \quad \times \Big( \| (\nabla_y \psi^{(n)}) (x, y, X_{n-2}) \|_{L_{x, y, X_{n-2}}^2} \| V(x-y) \varphi_t (x) \nabla \varphi_t (y) \|_{L_{x, y}^2} \| \psi^{(n-2)} (X_{n-2}) \|_{L_{X_{n-2}}^2} \nonumber \\
& \qquad + \| (\nabla_x \psi^{(n)}) (x, y, X_{n-2}) \|_{L_{x, y, X_{n-2}}^2} \| V(x-y) \varphi_t (x) \nabla \varphi_t (y) \|_{L_{x, y}^2} \| \psi^{(n-2)} (X_{n-2}) \|_{L_{X_{n-2}}^2} \nonumber \\
& \qquad + \| V(x-y) \psi^{(n)} (x, y, X_{n-2}) \|_{L_{x, y, X_{n-2}}^2} \| \nabla \varphi_t (x) \nabla \varphi_t (y) \|_{L_{x, y}^2} \| \psi^{(n-2)} (X_{n-2}) \|_{L_{X_{n-2}}^2} \Big). \nonumber
\end{align}

When $V^2 \leq D(1-\Delta)$,
\begin{align}
& \| V(x-y) \psi^{(n)} (x, y, X_{n-2}) \|_{L_{x, y, X_{n-2}}^2} \nonumber \\
&\leq D \left( \| (\nabla_x \psi^{(n)}) (x, y, X_{n-2}) \|_{L_{x, y, X_{n-2}}^2} + \| \psi^{(n)} (x, y, X_{n-2}) \|_{L_{x, y, X_{n-2}}^2} \right). \label{H2 derivative linear term 2}
\end{align}
Hence, from \eqref{H2 derivative linear term 1} and \eqref{H2 derivative linear term 2}, we have
\begin{align}
& \Big| \iint dx dy \left \langle a_x a_y \psi, V(x-y) \varphi_t (x) (-\Delta \varphi_t) (y) \psi \right \rangle \Big| \nonumber \\
&\leq C \sum_{n=2}^{\infty} \sqrt{n(n-1)} \| (\nabla_x \psi^{(n)}) (x, y, X_{n-2}) \|_{L_{x, y, X_{n-2}}^2} \| \psi^{(n-2)} (X_{n-2}) \|_{L_{X_{n-2}}^2} \label{H2 derivative linear estimate} \\
& \quad + C \sum_{n=2}^{\infty} \sqrt{n(n-1)} \| \psi^{(n)} (x, y, X_{n-2}) \|_{L_{x, y, X_{n-2}}^2} \| \psi^{(n-2)} (X_{n-2}) \|_{L_{X_{n-2}}^2} \nonumber \\
&\leq C \langle \psi, (H_0 + \N+1) \psi \rangle. \nonumber
\end{align}
Now, \eqref{H2 derivative expansion}, \eqref{H2 derivative nonlinear estimate}, and \eqref{H2 derivative linear estimate} show that
\begin{align}
\Big| \iint dx dy \langle \psi, V(x-y) \varphi_t (x) \dot{\varphi_t}(y) a_x^{\dag} a_y^{\dag} \psi \rangle \Big| \leq C \langle \psi, (H_0 + C (\N+1) ) \psi \rangle.
\end{align}

Other terms in the right hand side of \eqref{H2 derivative} can be estimated similarly. Hence, we get
\begin{align}
\langle \psi, \dot{H_2} (t) \psi \rangle \leq C \langle \psi, (H_0 + C (\N+1) ) \psi \rangle.
\end{align}
Furthermore, we also know that $H_2(t) - H_0 \geq -C (\N+1)$, or
\begin{equation}
H_2 (t) + C (\N+1) \geq H_0.
\end{equation}
(See Corollary 2.1 in \cite{GV2}.) Then, we can find constants $C_0$ and $C_1$ such that
\begin{equation}
\frac{d}{dt} \left( H_2 (t) + C_0 (\N+1) \right) \psi \rangle \leq C_1 \langle \psi, \left( H_2 (t) + C_0 (\N+1) \right) \psi \rangle.
\end{equation}
This proves the desired lemma.

\section*{Acknowledgment}
We are grateful to H.-T. Yau and Benjamin Schlein for helpful discussions.

\end{document}